\documentclass[%
 reprint,
 superscriptaddress,
 amsmath,amssymb,
 aps,
 prb,
longbibliography
]{revtex4-2}

\usepackage[final]{graphicx}
\usepackage{dcolumn}
\usepackage{bm}
\usepackage{xcolor}
\usepackage{color,soul}
\usepackage{algorithm}
\usepackage{algpseudocode}
\usepackage{url} 
\usepackage{makecell}
\usepackage{multirow}
\usepackage{booktabs}
\usepackage{siunitx}
\DeclareSIUnit\atom{atom}

\usepackage{orcidlink}
\usepackage{subfig}
\usepackage{changes}
\usepackage[normalem]{ulem}
\usepackage{hyperref}
\hypersetup{colorlinks = true, allcolors = .}
\usepackage{cleveref}
\usepackage[frozencache,cachedir=.]{minted}

\newcommand{\PY}{{\texttt{\detokenize{pyiron}}}}
\newcommand{\PM}{{\texttt{\detokenize{pacemaker}}}}
\newcommand{\LA}{{\texttt{\detokenize{LAMMPS}}}}
\newcommand{\CA}{{\texttt{\detokenize{calphy}}}}
\newcommand{\RU}{{\texttt{\detokenize{RuNNer}}}}
\newcommand{\AT}{{\texttt{\detokenize{atomicrex}}}}
\newcommand{\PYC}{{\texttt{\detokenize{pycalphad}}}}

\graphicspath{{graphics/}}

\begin{document}

\title{From electrons to phase diagrams with classical and machine learning potentials: automated workflows for materials science with \PY{}}


\author{Sarath Menon \orcidlink{0000-0002-6776-1213}} 
\email[]{s.menon@mpie.de}
\affiliation{Max-Planck-Institut f\"ur Eisenforschung GmbH, 40237 Düsseldorf, Germany}

\author{Yury Lysogorskiy }
\affiliation{ICAMS, Ruhr-Universit\"at Bochum, 44801 Bochum, Germany}

\author{Alexander L. M. Knoll}
\affiliation{Lehrstuhl f\"ur Theoretische Chemie II, Ruhr-Universität Bochum, 44780 Bochum, Germany}
\affiliation{Research Center Chemical Sciences and Sustainability, Research Alliance Ruhr, 44780 Bochum, Germany}

\author{Niklas Leimeroth \orcidlink{0009-0005-3906-4751}}
\affiliation{Technische Universit\"at Darmstadt, Fachbereich Material und Geowissenschaften, Fachgebiet Materialmodellierung, 64287 Darmstadt, Germany}

\author{Marvin Poul \orcidlink{0000-0002-6029-8748}}
\affiliation{Max-Planck-Institut f\"ur Eisenforschung GmbH, 40237 Düsseldorf, Germany}

\author{Minaam Qamar \orcidlink{0000-0002-3342-4307}}
\affiliation{ICAMS, Ruhr-Universit\"at Bochum, 44801 Bochum, Germany}

\author{Jan Janssen \orcidlink{0000-0001-9948-7119}}
\affiliation{Max-Planck-Institut f\"ur Eisenforschung GmbH, 40237 Düsseldorf, Germany}

\author{Matous Mrovec \orcidlink{0000-0001-8216-2254}}
\affiliation{ICAMS, Ruhr-Universit\"at Bochum, 44801 Bochum, Germany}

\author{Jochen Rohrer \orcidlink{0000-0002-4492-3371}}
\affiliation{Technische Universit\"at Darmstadt, Fachbereich Material und Geowissenschaften, Fachgebiet Materialmodellierung, 64287 Darmstadt, Germany}

\author{Karsten Albe \orcidlink{0000-0003-4669-8056}}
\affiliation{Technische Universit\"at Darmstadt, Fachbereich Material und Geowissenschaften, Fachgebiet Materialmodellierung, 64287 Darmstadt, Germany}

\author{Jörg Behler \orcidlink{0000-0002-1220-1542}}
\affiliation{Lehrstuhl f\"ur Theoretische Chemie II, Ruhr-Universität Bochum, 44780 Bochum, Germany}
\affiliation{Research Center Chemical Sciences and Sustainability, Research Alliance Ruhr, 44780 Bochum, Germany}

\author{Ralf Drautz \orcidlink{0000-0001-7101-8804}}
\affiliation{ICAMS, Ruhr-Universit\"at Bochum, 44801 Bochum, Germany}

\author{J\"org Neugebauer \orcidlink{0000-0002-7903-2472}}
\email[]{neugebauer@mpie.de}
\affiliation{Max-Planck-Institut f\"ur Eisenforschung GmbH, 40237 Düsseldorf, Germany}

\date{\today}

\begin{abstract}
We present a comprehensive and user-friendly framework  built upon the \PY{} integrated development environment (IDE), enabling researchers to perform the entire Machine Learning Potential (MLP) development cycle consisting of (i) creating systematic DFT databases, (ii) fitting the Density Functional Theory (DFT) data to empirical potentials or MLPs, and (iii) validating the potentials in a largely automatic approach. The power and performance of this framework are demonstrated for three conceptually very different classes of interatomic potentials: an empirical potential (embedded atom method - EAM), neural networks (high-dimensional neural network potentials - HDNNP) and expansions in basis sets (atomic cluster expansion - ACE). As an advanced example for validation and application, we show the computation of a binary  composition-temperature phase diagram for Al-Li, a technologically important lightweight alloy system with applications in the aerospace industry.   
\end{abstract}

\maketitle

\section{Introduction}

The advent of machine learning interatomic potentials (MLPs) is revolutionising the field of computational materials science, enabling simulations of large systems and complex material properties with \textit{ab initio} accuracy \cite{Behler2016,Deringer2019,Unke2021,Friederich2021,Behler2021}.
However, the development of these data-driven interatomic potentials is a computationally intensive task that needs automated and reliable workflows.

The life cycle of MLP development can be broadly divided into the following tasks: (i) generating a database containing reference data, (ii) fitting the model parameters to the reference data, and (iii) validating the resulting parametrization for a specified range of properties.
Furthermore, it is often necessary to provide a feedback loop between the tasks via an active learning approach to ascertain transferability~\cite{Lysogorskiy2021,Tokita2023}.

The initial task of setting up the reference database usually requires to perform many thousands of density functional theory (DFT) calculations for a broad range of atomic environments that span the configuration space of interest as completely as possible. Such computations can nowadays be facilitated using either general workflow frameworks \cite{Huber2020,Mathew2017,Gjerding2021,Janssen2019} or tools designed specifically for a particular MLP class \cite{Duff2023,Zeng2023,Rohskopf2023,Vandenhaute2023,Gelzinyte2023,Wen2022}. 
Nevertheless, there is still lack of standardized workflow setups, computational metaparameters and structural databases. Therefore, each research group relies mostly on their own expertise and experience. This may not only lead to inconsistencies in the generated data (for instance, due to variations in DFT settings, such as the exchange-correlation functional, Brillouin zone sampling or plane wave cutoff)~\cite{kratzer2019basics, bosoni2023verify}, 
but it can also strongly limit exchange of data from different sources and their collection into greater databases. 

\begin{figure*}[ht!]
\centering
\includegraphics[width=1.0\textwidth]{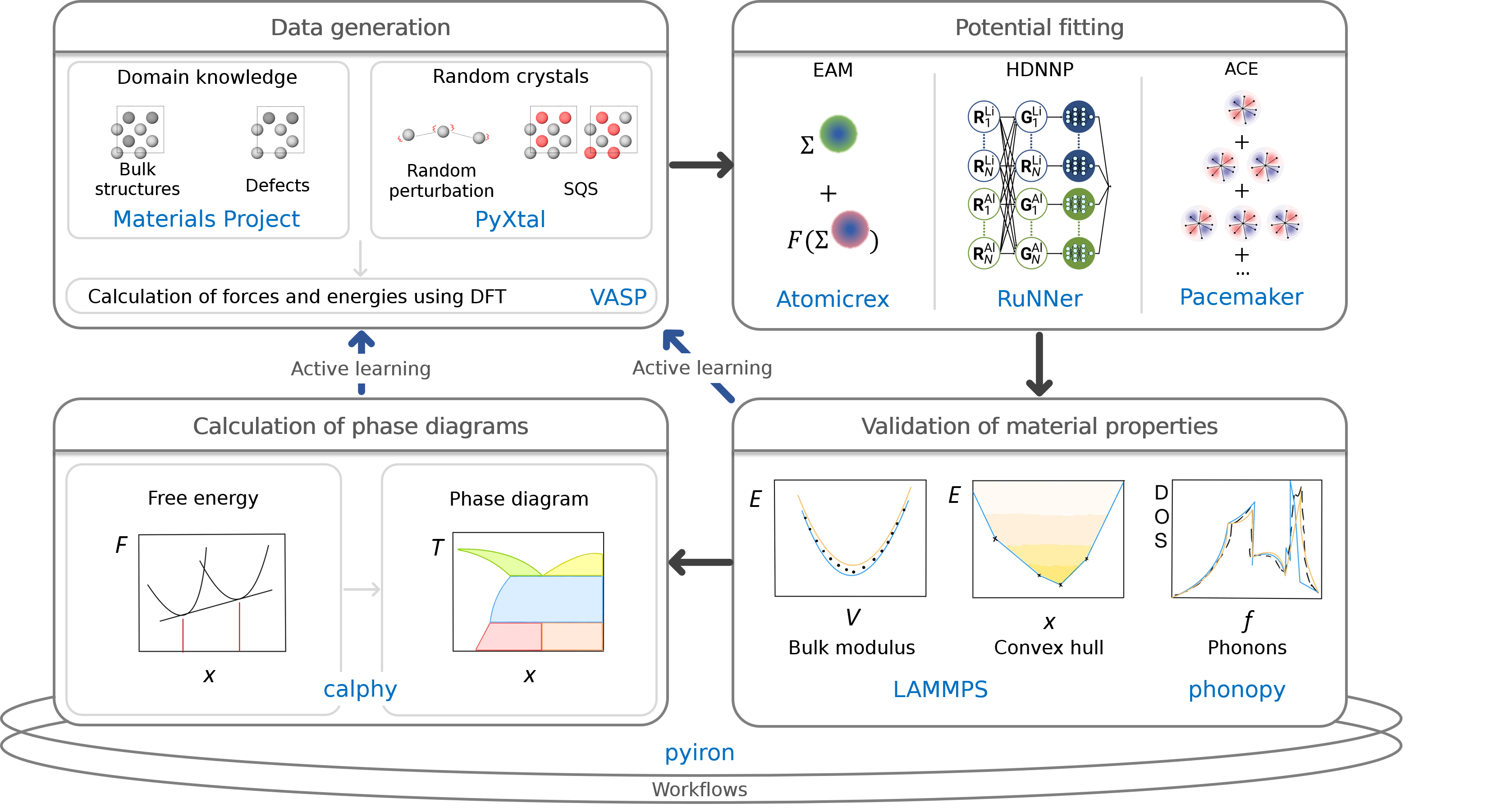}
\caption{A schematic illustration of the MLP development cycle. Here, \PY{} is employed as a workflow manager to combine different software tools and packages.} 
\label{fig:layout}
\end{figure*}

The situation remains similar when it comes to the second stage of MLP development, namely, fitting the model parameters. Many optimization algorithms and software tools are  tailored to a particular class of potentials and are not easily transferable. 
Thus, a researcher must not only identify an appropriate type of potential that is suited for the system of interest, but often needs to learn a variety of specific software tools each of which uses its own terminology. 

The final task in the development cycle is a thorough validation of the fitted parametrization. It should be stressed that simple correlations between the predictions of the potential and the reference DFT data, e.g. for energies and forces, are in most cases not sufficient and may be even misleading. 
It is crucial to evaluate not only fundamental physical properties, such as energy-volume curves, elastic moduli or phonon spectra, but to perform also dynamical simulations at finite temperatures that scrutinize spurious behavior of the model outside of the training domain. Existing initiatives~\cite{Becker2013, Hale2018, Lysogorskiy2019} have mostly focused on classical interatomic potentials while automatized validations of MLPs are still rare~\cite{P5893}. This makes it difficult for users to determine a safe application range of a particular MLP parametrization for their application. A thorough validation is also indispensable before applying the model in simulations of complex material properties, such as studies of extended defects, phase transformations, or predictions of phase diagrams.


Our aim is to demonstrate the whole MLP development cycle for three representative model potentials
to elucidate the complete process to interested researchers, not on a benchmark comparison of different types of potentials. We introduce a set of standardised workflows that cover all aspects from generation of DFT data to MLP fitting and validation, as schematically illustrated in Fig. \ref{fig:layout}.
As an example of an advanced application, we evaluate the phase diagrams for a prototypical binary system. The workflows presented here are reusable, reproducible, and most importantly, largely automated. While exposing all intricacies of the methods involved, we show that they significantly reduce the technical complexity. By providing the computer codes and software tools, we encourage to use this manuscript as a practitioner's guide into the field of modern MLP development as well as advanced thermodynamic applications. 

As a model case, we chose the binary Al-Li system \cite{Abd2018, Hallstedt2007}. Al-Li alloys are well suited for aerospace applications since they exhibit low density and high mechanical strength \cite{Gupta2006, Rioja1998}. Apart from a recent work \cite{Liu2024}, there is a lack of interatomic potential for this system, making it both desirable and a challenging option from the perspective of MLP development. We selected three prototypical examples of interatomic potentials: a classical central-force potential based on the embedded atom method (EAM)  \cite{dawEmbeddedatomMethodDerivation1984, baskesApplicationEmbeddedAtomMethod1987}, a high-dimensional neural network potential (HDNNP) \cite{Behler2007,Behler2021b}, and the atomic cluster expansion (ACE) \cite{Drautz2019, Lysogorskiy2021}.
We use \texttt{calphy} \cite{Menon2021} for the calculation of phase diagram, and employ \PY{} \cite{Janssen2019} as a workflow creation and management environment to bring together various software tools. Our goal is to enable seamless creation and validation of interatomic potentials while taking a step towards the FAIR (Findable, Accessible, Interoperable, and Reusable) data and software principles \cite{Wilkinson2016, Chue2021} in the field of MLP development.

\section{Results}

\subsection{\PY{} as a platform for automated workflows}\label{sec:pyiron}

We employ \PY{}\cite{Janssen2019} as a workflow environment for all stages of the MLP development and validation, as illustrated in Fig. \ref{fig:layout}. The development
cycle is facilitated with \PY{} by connecting the fundamental building blocks:

\begin{enumerate}
    \item \label{enum:blocks_a} Generic and easy-to-use structure generation tools that combine standard software libraries in the field of computational materials science such as \texttt{ASE} \cite{Larsen2017}, \texttt{pymatgen} \cite{Ong2013}, \texttt{PyXtal} \cite{pyxtal}, and others in a convenient and interoperable package;
    \item \label{enum:blocks_b} An interoperable interface to a variety of electronic structure and atomistic simulation software packages such as \texttt{VASP} \cite{kresse1993ab,kresse1996efficiency,kresse1996efficient} and \texttt{LAMMPS} \cite{Plimpton1995};
    \item \label{enum:blocks_c} A common storage format for energies, forces and stresses that can be used efficiently for hundreds of thousands of training configurations implemented in \PY{} as the class \texttt{TrainingContainer} (see Sec. \ref{sec:tc});
    \item \label{enum:blocks_d} A common interface to the fitting tools used in this work, namely, \PM{} \cite{Lysogorskiy2019, Bochkarev2022}, \RU{} \cite{Behler2015, Behler2017, runnerase2021}, and \AT{} \cite{stukowskiAtomicrexGeneralPurpose2017}, implemented in \PY{} as the \texttt{PotentialFit} class;
    \item \label{enum:blocks_e} A common interface to validation workflows and tools for thermodynamic properties such as \texttt{phonopy} \cite{phonopy-phono3py-JPCM} and \CA{} \cite{Menon2021}.
\end{enumerate}

Block (\ref{enum:blocks_a}) enables users with different backgrounds to generate easily new configurations or to import them from existing databases. The common interface in Block (\ref{enum:blocks_b}) provides a seamless switching between different quantum engines or simulation protocols to create training data with minimal changes in an existing workflow. It can also be employed to design and test a workflow using a lower-level method, which is computationally cheap, before switching to a production run using a higher-level theory. 
Blocks (\ref{enum:blocks_c}) and (\ref{enum:blocks_d}) provide flexibility to experiment with different MLP formalisms on the same data or selected subsets.  In addition, Block (\ref{enum:blocks_c}) 
provides analysis and plotting routines for all types of training data generated in Block (\ref{enum:blocks_b}).
Finally, Block (\ref{enum:blocks_e}) provides access to validation routines, helping a user to assess the quality of the fitted MLPs.

\subsection{Construction of a reference DFT dataset for the Al-Li system}

Selection of the reference data needed to parametrize an interatomic potential is one of the most important steps in the life cycle of MLP development.
For creation of the reference DFT data we employed VASP\,5.4~\cite{kresse1993ab,kresse1996efficiency,kresse1996efficient}, using workflows as described in Sec. \ref{sec:vasp_workflow}. 
We employed the projector augmented wave (PAW) method~\cite{kresse1999ultrasoft} and the GGA-PBE exchange correlation functional~\cite{perdew1996generalized}. 
Several convergence tests were conducted to ensure the obtained energies and forces are highly accurate and consistent. 
These tests were carried out for three representative structures, namely, face-centered cubic (fcc) Al, body-centered cubic (bcc) Li, and the B32-type $\beta$-LiAl, in a range of 30\,\% volumetric strain around their respective equilibrium volumes.
Based on these tests, the following DFT settings were used: a plane wave cutoff of 750\,eV, a k-mesh spacing of 0.1\,\AA$^{-1}$, and the Fermi smearing method with a width of 0.1\,eV.  With these settings we observed less than 0.5 meV/atom difference in the energies as compared to calculations performed at 800\,eV plane wave cutoff and 0.05\,\AA$^{-1}$ k-mesh spacing.

There exist multiple strategies for generation of relevant atomic configurations. We find that a combination of domain knowledge, active learning, and random search can be employed effectively for the construction of a balanced training dataset. In this three-step strategy, domain knowledge is employed first to select structures based on common structural prototypes available in standard crystallographic databases~\cite{Jain2013}. 
Thereafter, we use active learning algorithms during validation and simulations, and random configurations obtained using a random-structure-search procedure~\cite{poul2023systematic} to augment the dataset.

The domain-knowledge step is focused on structures that are known to be important for the system of interest and to ensure that they are represented with high accuracy. In this work, we queried both elemental and binary structures with formation energies less than or equal to zero from the Materials Project database~\cite{Jain2013}. 
Subsequently, a series of transformations was applied to these structures and their supercells, including uniform and non-uniform deformations of the cells and random displacements of atoms (see Supplementary information for details). 
These steps ensure that not only perfect bulk structures but also their distortions, which are crucial to reproduce elastic and vibrational properties, are included in the training data. 

We then used active learning to ensure that even during extended simulations, which are needed to compute thermodynamic properties (See Sec. \ref{sec:calphy}), potentials remain stable and accurate.
Within the active learning loop, we iteratively selected structures based on high uncertainty indicators~\cite{lysogorskiy2023active} derived from running molecular dynamics simulations for several Al-Li phases of interest.

Finally, we added random structures that are far from equilibrium. This step ensures a broader coverage of the configurational space.
Relying on domain knowledge and active learning only may result in a short sighted training set that hampers the extrapolation capabilities of the fitted MLPs, whereas a random sampling-based approach alone might lead to the risk of missing or under-representing important phases in a material.
The random structures were generated following a recent workflow~\cite{poul2023systematic} that utilizes the PyXtal~\cite{fredericks2021pyxtal} software as described in Sec. \ref{sec:pyxtal_workflow}. 
We generated an initial set of structures of each space group for varying numbers of atoms ( Not all space groups can be generated for every composition of the cells due to Wyckoff multiplicity constraints.),
which are then relaxed using DFT first allowing only the volume to vary, followed by a full relaxation of the cell shape and internal degrees of freedom. 
It is not necessary that these relaxations lead to highly accurate minima in the potential energy landscape, so low accuracy DFT calculations can be employed to speed up this step.
These relaxed structures are then recomputed using the required precision to ensure consistency of the training dataset.
This procedure, very similar to {\it ab initio} structure search methods~\cite{pickard2011ab}, and originally developed for this purpose, has recently been applied for machine learning potentials ~\cite{yanxon2020neural}. 
The primary advantage of this approach is that it can help to find basins of the potential energy surface without domain knowledge, while also exposing the potential to a greater variety of structural and chemical environments \cite{poul2023systematic}.

Finally, random displacements and variations of the cell shape and size were applied to the relaxed structures to obtain additional samples around the minima of the potential energy surface.
Detailed parameters for the random perturbations are described in the
supplementary information.
The initial set of random crystals as well as structures resulting from both the minimization steps and the random perturbations are then combined and added to the training set.
The complete workflow is implemented in \PY{} and the primitives introduced in the previous section\,\ref{sec:pyiron}.

Through the combination of these three strategies --- domain-knowledge, active learning, and random search --- we were able to construct a robust and extensive atomic structure data set that captures a wide range of configurations.
The distributions of DFT reference data over energy, volume and composition are
shown in Fig.~\ref{fig:dft_evc_distr} and further information is provided in
the supplementary information.

\begin{figure}
    \centering
     \includegraphics[width=0.5\textwidth]{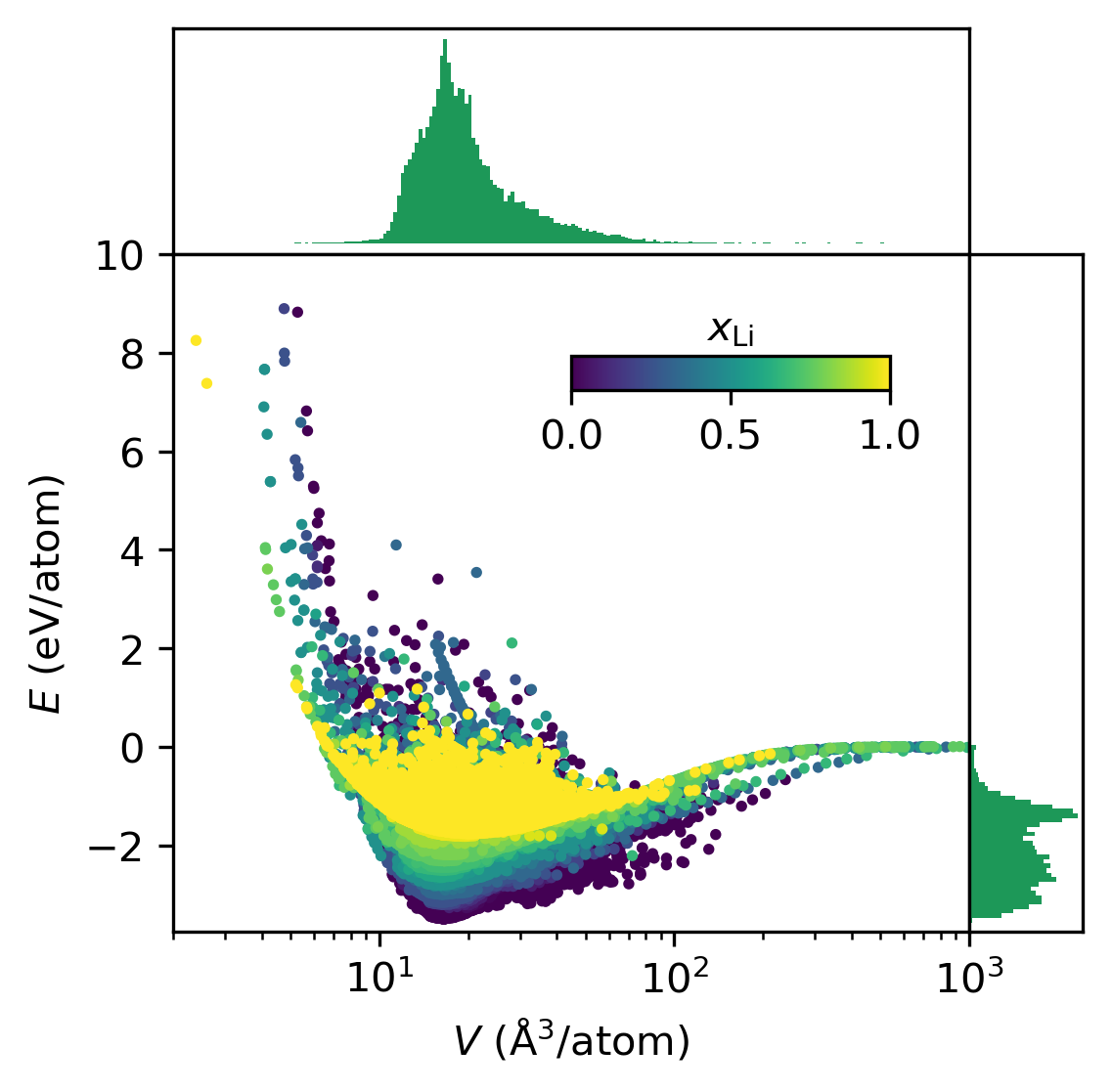}
    \caption{Energy distribution of the DFT reference data set as a function of the atomic volume. The fraction of lithium atoms in each structure is represented by the color of the points. }%
    \label{fig:dft_evc_distr}%
\end{figure}

\subsection{Training of data-driven interatomic potentials}

The training of MLPs is usually carried out using dedicated computer codes that are tailored to a particular model architecture. In our case, we used \texttt{atomicrex}~\cite{stukowskiAtomicrexGeneralPurpose2017}, \texttt{RuNNer}~\cite{Behler2015, Behler2017,runnerase2021} and \texttt{pacemaker}~\cite{Bochkarev2022} to fit the EAM, HDNNP and ACE parametrizations, respectively. 
Due to differences in the fitting procedures, the training data sets needed to be adjusted to the respective codes and models. 

\subsubsection{EAM}

For the EAM potential, we started by fitting potentials for the single elements,
following an approach outlined by Mishin et al.~\cite{mishinStructuralStabilityLattice2001}.
This approach guarantees an exact fit of the lattice constant,
cohesive energy and bulk modulus by constraining the fitting parameters accordingly.
Parameters of functions describing Al-Li were fitted
while keeping the single element parameters constant.
The training data was limited to the domain knowledge subgroup of the whole set, containing 2081 structures.
Because the potential has less than 100 adjustable parameters,
this amount of training data is sufficient, and leads to faster parametrization routines.
Furthermore, the functional form of EAMs have limited flexibility,
so training to randomly sampled structures far from equilibrium could impact the accuracy of the more important low-energy structures.
In the fitting process, energies were weighted based on the distance of the formation energy to the convex hull $E_{D}$ (in \si{eV/atom}) as
\begin{equation}
    W_E = \frac{100}{(E_D+0.2)^4},
\end{equation}
while a uniform weight of one was applied for forces. Further details about the fitting procedure are provided in the Methods section and in Ref.~\cite{mishinStructuralStabilityLattice2001}.

\subsubsection{HDNNP}

Before starting the HDNNP training process, the training data set was refined by eliminating structures which were not relevant for the material properties of interest. These included structures containing isolated atoms without any bonding partners within a radius of \SI{12}{\angstrom}, structures with large positive formation energies or highly repulsive force components, and all structures with atomic volume outside the interval of \SI{10}{\angstrom\cubed}/atom - \SI{50}{\angstrom\cubed}/atom. In total, 4915 data points were removed. \par

Using \textsc{runnerase} \cite{runnerase2021}, we then carried out a grid search to optimize the hyperparameters required for the HDNNP training performed with the \texttt{RuNNer} code \cite{Behler2015, Behler2017}. In particular, this includes the number, short-range cutoff radius and parameters of the atom-centered symmetry functions (ACSFs) \cite{Behler2011} describing the atomic environments, the hyperparameters of the optimisation algorithm (Kalman filter \cite{Kalman1960}), and the neural network architecture. For each trial, HDNNPs were trained on five randomly selected mini-batches of 200 data points and three random initialization seeds each. \SI{1}{\pico\second} $NVT$ MD rapid heating runs between \SI{300}{\kelvin} and \SI{1000}{\kelvin} were performed to test the capabilities of each potential in a basic application. Finally, the best hyperparameters were selected based on the training accuracy and simulation stability (length of the stable trajectories, number of extrapolations) which were achieved across the 15 members of the group. The selected hyperparameters are presented in supplementary material, Tab.\,III. Then, a HDNNP was trained with the selected settings on the entire training dataset, randomly separated into a training (90\%) and testing set (10\%). All data points have been equally weighted in the training process. The same applies to the relative weight of total energies and force components. \par

\begin{figure*}
    \centering
     \includegraphics[width=0.9\linewidth]{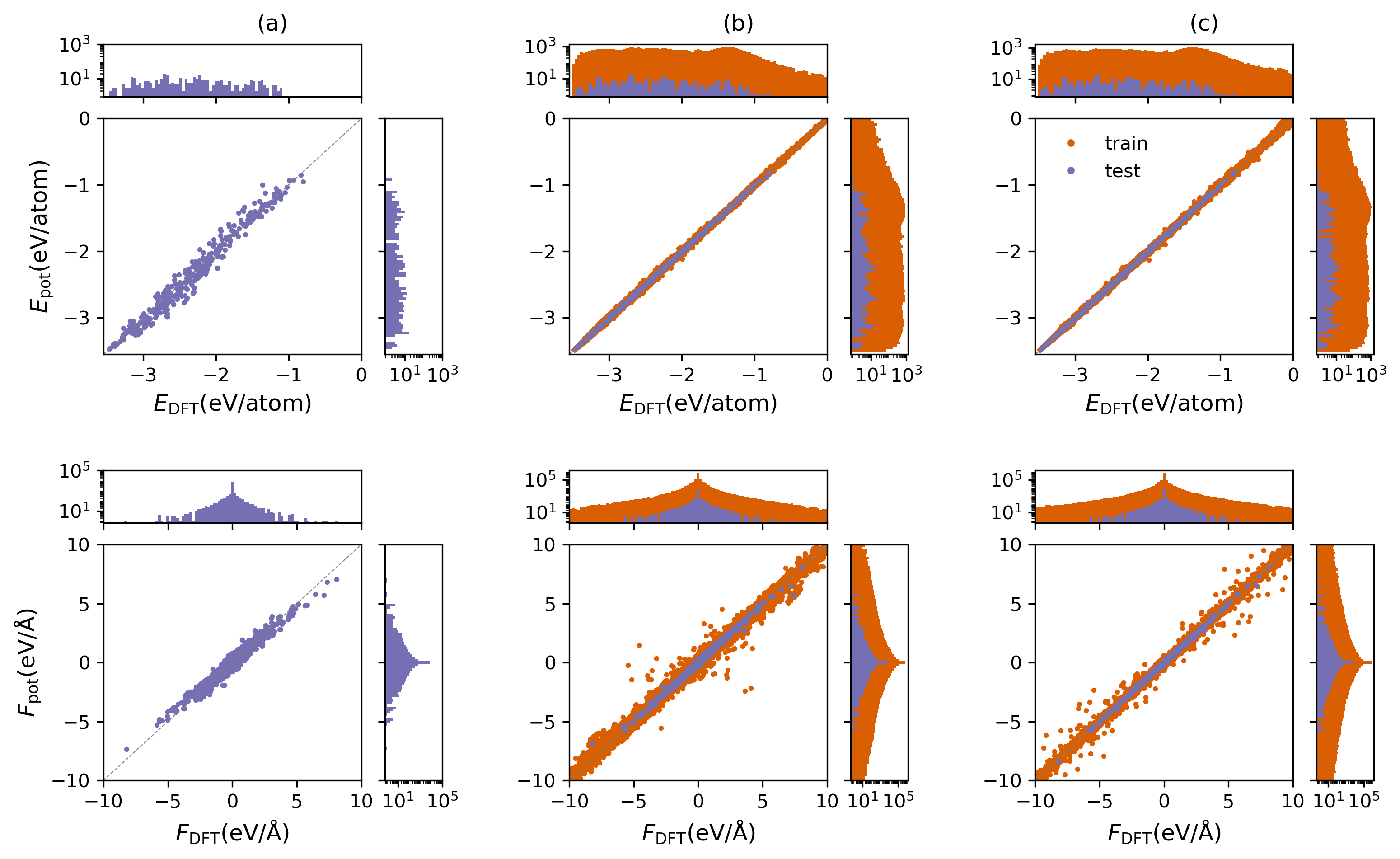}
    \caption{Energy and forces predicted by the potentials ($E_{\mathrm{pot}}$ and $F_{\mathrm{pot}}$) compared to DFT data for (a) EAM, (b) HDNNP, and (c) ACE. The corresponding values of the RMSE and MAE are given in Table~\ref{tab:comparison_potentials}. Note that training data (purple) is a different subset of the reference DFT dataset for each potential, while test data (orange) is based on a common test set including only structures within 1~eV/atom above the convex hull of the system.}
    \label{fig:test_train}
\end{figure*}

\begin{table*}[]
    \centering
    \caption{Energy and force RMSE (MAE) of EAM, HDNNP, and ACE with respect to DFT. Test set metrics are given for a common test set including only structures within 1~eV/atom above the convex hull of the system. The size of the dataset and the energy, force and volume ranges of training and testing data are also summarized.
    }
    \begin{tabular}{ccccccccc}
    \toprule
    \multirow{2}{*}{Potential} &
    \multirow{2}{*}{Dataset Size} &
    Energy Range &
    Force Range &
    Volume Range &
    \multicolumn{2}{c}{$E$ RMSE (MAE)} &
    \multicolumn{2}{c}{$F$ RMSE (MAE)}
    \\
    &
    &
    [\si{\eV\per\atom}] &
    [\si{\eV\per\angstrom}] &
    [\si{\angstrom\cubed\per\atom}] &
    \multicolumn{2}{c}{[\si{\milli\eV\per\atom}]} &
    \multicolumn{2}{c}{[\si{\milli\eV\per\angstrom}]}
    \\[.2cm]
    \multicolumn{5}{c}{} &
    Train &
    Test &
    Train &
    Test
    \\
    \midrule
    EAM &
    2081 &
    58 &
    52 &
    235 &
    554 (82.2) &
    118 (89.9) &
    192 (116) &
    143 (68.7)
    \\
    HDNNP &
    50834 &
    3.5 &
    20 &
    40 &
    10.6 (7.1) &
    10.2 (6.9) &
    64.2 (30.6) &
    49.2 (21.0)
    \\
    ACE &
    51082 &
    50 &
    40 &
    600 &
    12.2 (7.5) &
    9.6 (6.6)&
    41.4 (16.9)&
    26.5 (11.7)
    \\
    \bottomrule
    \end{tabular}
    \label{tab:comparison_potentials}
\end{table*}

\subsubsection{ACE}

The ACE parameterization was carried out using the \texttt{Pacemaker} package~\cite{Bochkarev2022}. A cutoff for all interactions was set to 7\,\AA, based on the range of DFT interactions.
The total number of basis functions per element was set to 1000.
The resulting maximum radial and angular indices, dependent on the correlation order, as well as other ACE parameters are presented in Table~II in supplementary material.

The total dataset was randomly divided into training and testing sets with a ratio of 95\% to 5\%.
Weights for the training structures were assigned based on their energy distance to the convex hull, following the energy-based weighting method~\cite{Bochkarev2022}.

The fitting was performed in two stages. In the first stage, a higher emphasis was placed on forces ($\kappa=0.99$), while in the second stage a more balanced distribution of energy-forces weights ($\kappa=0.3$) was used.

A strong core repulsion pair potential was added at distances below 2\,\AA.


\subsubsection{Comparison of training outcomes}

An overview of all training datasets as well as the achieved training accuracy for all three potentials is given in Table~\ref{tab:comparison_potentials} and the correlations between predicted and reference values for energies and force components are provided in Fig.\,\ref{fig:test_train}.


It is seen that all three fitting methods yield favorable training results despite the different train set sizes and compositions. In line with our expectations, the physically-inspired EAM requires the least amount of training data spanning over large energy, force and volume ranges, albeit at the cost of higher training errors. In contrast, the HDNNP and ACE utilize most of the available training data and reach smaller errors with respect to the DFT reference values than EAM. Due to the higher weighting of low-energy structures employed in the training of the ACE potential, there are less outliers around forces close to zero in Fig.\,\ref{fig:test_train} (c), while such a weighting has not been applied in the HDNNP training.

To facilitate the validation and to compare objectively the accuracy of all potentials, we created a single test dataset containing only structures that were not part of any training dataset. The test structures were restricted to lie within \SI{1}{\eV}/atom or less above the convex hull, as these represent the physically most relevant subset for the phase diagram simulations. In Table~\ref{tab:comparison_potentials} and Fig.\,\ref{fig:test_train}, we depict test set metrics for this common test dataset only.
For all potentials, the test error metrics are smaller than those for the training datasets. In the case of ACE, the overall metrics are additionally biased due to the non-uniform distribution of energy-based weights.

\subsection{Validation approach and strategy}


Once the potentials have been parameterized with the desired accuracy, they must be extensively validated. Energy and force RMSEs of the final fit provide a first quantitative assessment of the potentials with respect to the reference data. However, it is mandatory to evaluate a broader range of fundamental material properties and to compare them to DFT reference data, and when applicable, experimental observations.

\begin{figure*}
    \centering
     \includegraphics[width=0.95\textwidth]{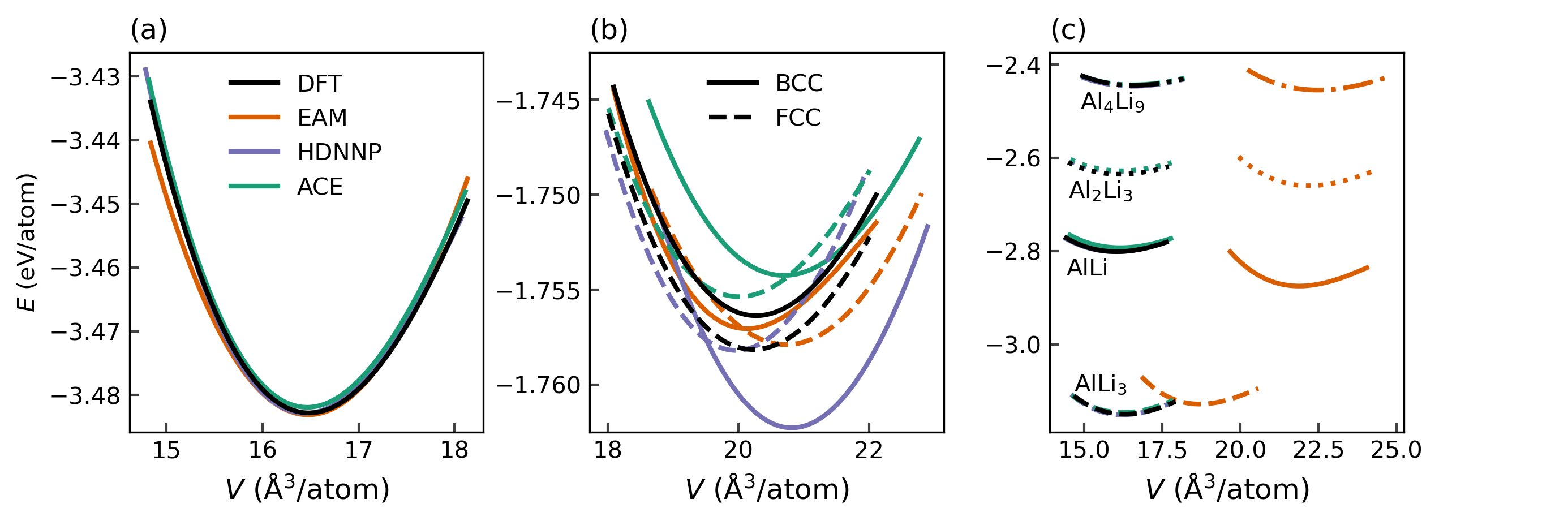} 
    \caption{Equation of state curves for (a) pure fcc Al, (b) pure bcc and fcc Li, and (c) different AlLi compounds as predicted by the EAM, HDNNP, and ACE potentials. The DFT reference is shown in black. In (b), as the bulk modulus of Li (11 GPa for bcc and 13 GPa for fcc) is lower than the respective value of Al (76 GPa), the range of energies is small, approximately 0.02 eV/atom, resulting in rather large visual discrepancies. For the binary compounds in (c), the predictions of the HDNNP, ACE, and DFT essentially coincide, and are thus hardly distinguishable. }
    \label{fig:eV_curves}%
\end{figure*}

An elementary assessment of transferability is to compare energies of important bulk phases as a function of atomic volume. This data is fitted to the Murnaghan equation of state to obtain Murnaghan curves, that contain not only valuable information about the mutual stability of various phases, but also can be used for an estimation of their bulk moduli. Fig.~\ref{fig:eV_curves} shows the Murnaghan curves as predicted by all three potentials for the fcc phase of Al, bcc and fcc phases of Li, and four binary phases (AlLi, Al$_3$Li, Al$_4$Li$_9$, and Al$_2$Li$_3$) that appear in the phase diagram \cite{Hallstedt2007}.

All potentials agree well with DFT for the ground states of Al and Li, with some minor variations observed for both Li phases of the order of a few meV (Note that the ground state according to reference PBE-DFT calculations is fcc, albeit with a small energy difference compared to bcc, as seen in Fig. \ref{fig:eV_curves} (b) ). When considering the binary phases in Fig.~\ref{fig:eV_curves} (c), a clear distinction is observed between the EAM potential and the MLPs. While the MLPs predict both the atomic volumes as well as the formation energies in excellent agreement with DFT, the EAM potential shows a considerable overestimation of the atomic volumes. 

\begin{figure*}
\centering
\includegraphics[width=1.\textwidth]{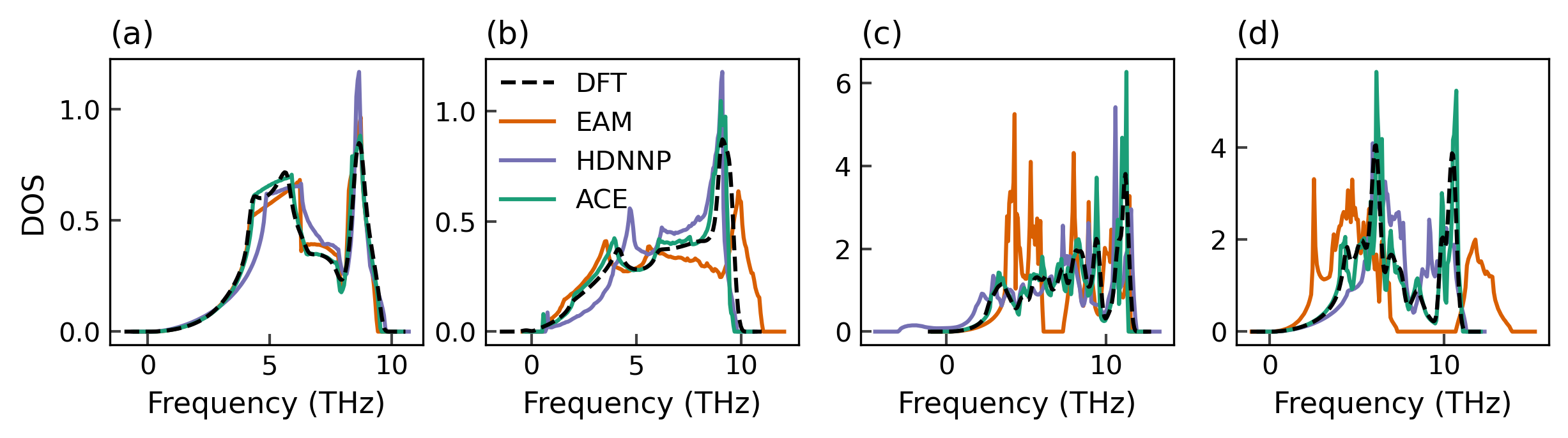}
\caption{Phonon density of states for (a) fcc Al, (b) bcc Li, (c) AlLi, and (d) Al$_3$Li, as predicted by EAM, HDNNP, and ACE in comparison with the DFT reference.}
\label{fig:pDOS}
\end{figure*}

The phonons predicted by the potentials are related to vibrational properties of materials and reflect the model's behaviour for small perturbations near the equilibrium ground state structure that are relevant for an accurate reproduction of phase diagrams. Figure~\ref{fig:pDOS} shows the phonon densities of states for fcc Al, bcc Li, AlLi and Al$_3$Li. Note that we restrict ourselves to the two binary phases with $x_\mathrm{Li} \leq  0.5$, as this is the region considered in the phase diagram calculations (see Sec. \ref{sec:pd}). Similar to the Murnaghan curves, it is observed that HDNNP and ACE both predict the phonon DOS with good accuracy, while the EAM potential shows significant deviations for the binary phases.

\begin{figure}
\centering
\includegraphics[width=0.48\textwidth]{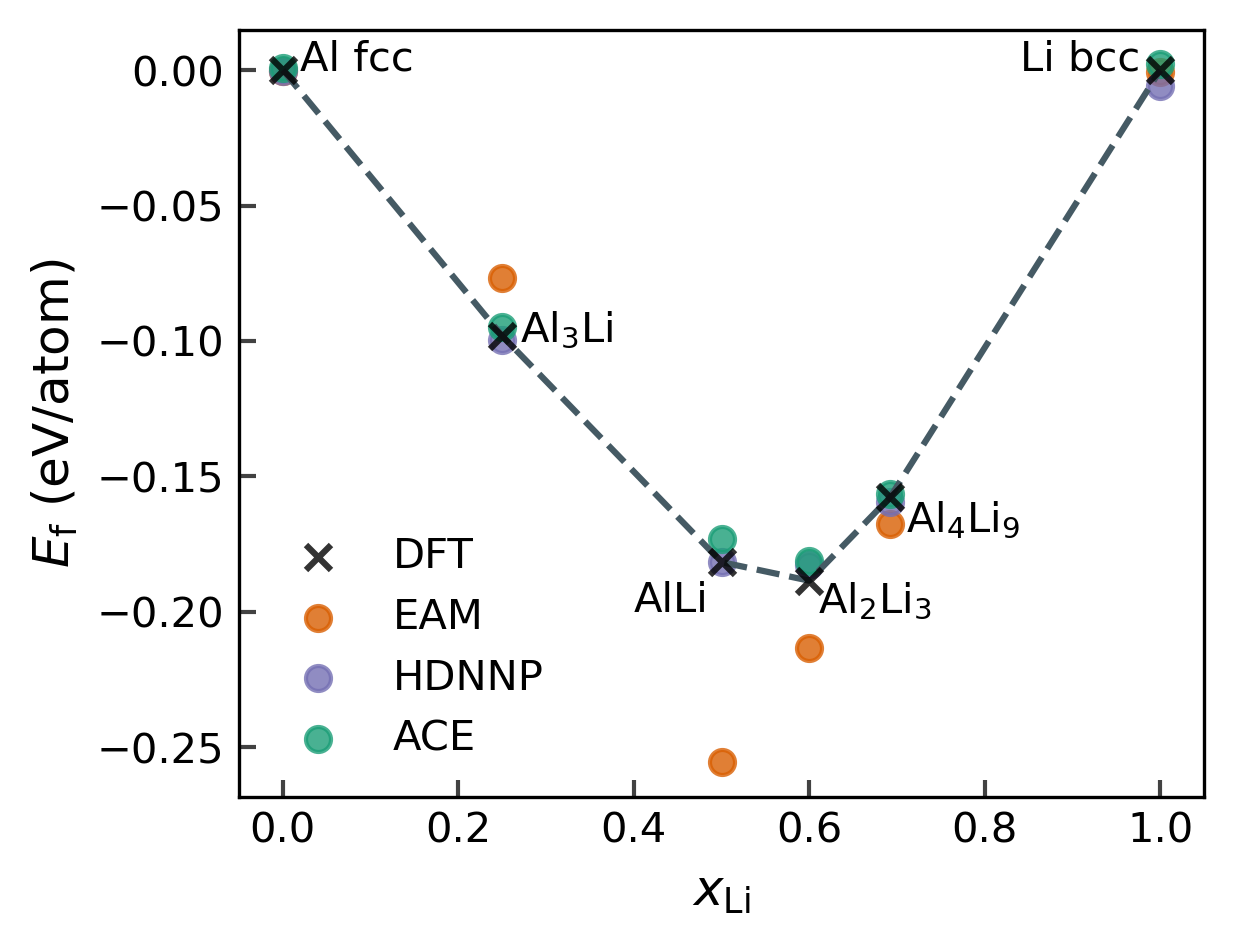}
\caption{The convex hull for the Al-Li binary system as predicted by EAM, HDNNP, and ACE. The black dashed line connects the points along the DFT convex hull.}
\label{fig:convexhull}
\end{figure}

As an initial step before evaluating the binary phase diagram, we evaluate and plot the formation energies of the binary phases as a function of Li content in Fig.~\ref{fig:convexhull}. The so-called convex hull can in fact be thought of as a binary phase diagram at zero Kelvin. The convex hull plot allows the formation energies to be directly compared to DFT and our calculations show that both HDNNP and ACE predict the formation energies well with DFT accuracy while EAM predictions deviate from the reference.

The elastic matrix elements C$_{11}$, C$_{12}$ and C$_{44}$ for the fcc Al and bcc Li ground states are computed with the fitted potentials and reported in Table~\ref{tab:elastic}. For fcc Al, the machine learning potentials match the DFT reference data very well while the EAM potential overestimates C$_{11}$ and underestimates the other two elastic constants. For bcc Li, all potentials provide a good description of the elastic matrix.


\begin{table}[]
\caption{Elastic constants of elemental aluminium and lithium, given in GPa, as predicted by the three potentials and the DFT reference method.}
\begin{tabular}{lcccc}
\hline
\multicolumn{5}{c}{Al-fcc}                      \\ \hline
\multicolumn{1}{c}{} & DFT & EAM & HDNNP & ACE \\ \hline
C11                  & 129 & 98  & 131    & 130 \\
C12                  & 52  & 67  & 67     & 67  \\
C44                  & 32  & 46  & 49     & 39  \\ \hline
\multicolumn{5}{c}{Li-bcc}                      \\ \hline
C11                  & 15  & 15  & 12     & 13  \\
C12                  & 13  & 14  & 13     & 12  \\
C44                  & 11  & 12  & 12     & 11  \\ \hline
\end{tabular} \label{tab:elastic}
\end{table}

\subsection{Construction of thermodynamic phase diagrams} \label{sec:pd}

Phase diagrams provide critical information about the material system, the phases that are predicted to be stable at the given thermodynamic conditions, and the conditions at which one phase transitions to another, or two phases coexist.
Phase diagrams are, therefore, a crucial and challenging test for interatomic potentials.
In general, the given interatomic potential should be able to reproduce the key aspects of the phase diagram, or at least parts of it, pertaining to the expected thermodynamic region where the interatomic potential is to be employed.\\

The CALPHAD method \cite{Kaufman1956} is perhaps the most well-known method for the calculation of phase diagrams, aided by experimental observations of thermodynamic properties of a system.
From an atomistic perspective, different methods exist for the determination of phase diagrams \cite{Vega2008, Chew2023}. 
Broadly, the methods either evaluate phase stability directly, through approaches such as coexisting phase simulations \cite{Opitz1974, Ladd1977, Frenkel1991}, or indirectly, by determining the Gibbs free energy or chemical potential of the relevant phases \cite{FrenkelSmit2002}. 
We follow the approach of calculating free energies, using the thermodynamic integration method, in which the free energy difference between a given system and a reference system is calculated \cite{Kirkwood1935, Smit2001}. 
We combine thermodynamic integration with non-equilibrium Hamiltonian interpolation and reversible scaling to obtain the free energies efficiently (see Methods for more details).
The workflow for such a calculation boils down to the code as described in Sec. \ref{sec:calphy}.
For this methodology, \textit{a priori} information about the relevant phases is needed, which is motivated by the currently established phase diagram \cite{Gayle1984, Hallstedt2007}.
In order to have a set of robust, automated, and efficient workflows for the phase diagram determination, we consider only substitutional defects in the off-stoichiometric compounds, and limit ourselves to the left side of the phase diagram, until $x_\mathrm{Li}=0.5$.
Therefore, we consider the fcc Al, AlLi in the bcc-like B32 lattice, and the liquid.
Furthermore, the L$1_2$ $\mathrm{Al}_3\mathrm{Li}$ appears as a metastable phase in the experimentally determined phase diagram \cite{Gayle1984}, and on the convex hull determined through DFT calculations (see Fig. \ref{fig:convexhull}), which makes it an interesting candidate to be considered in the calculation of the phase diagram. 

\begin{figure}[ht]
\centering
\includegraphics[width=0.48\textwidth]{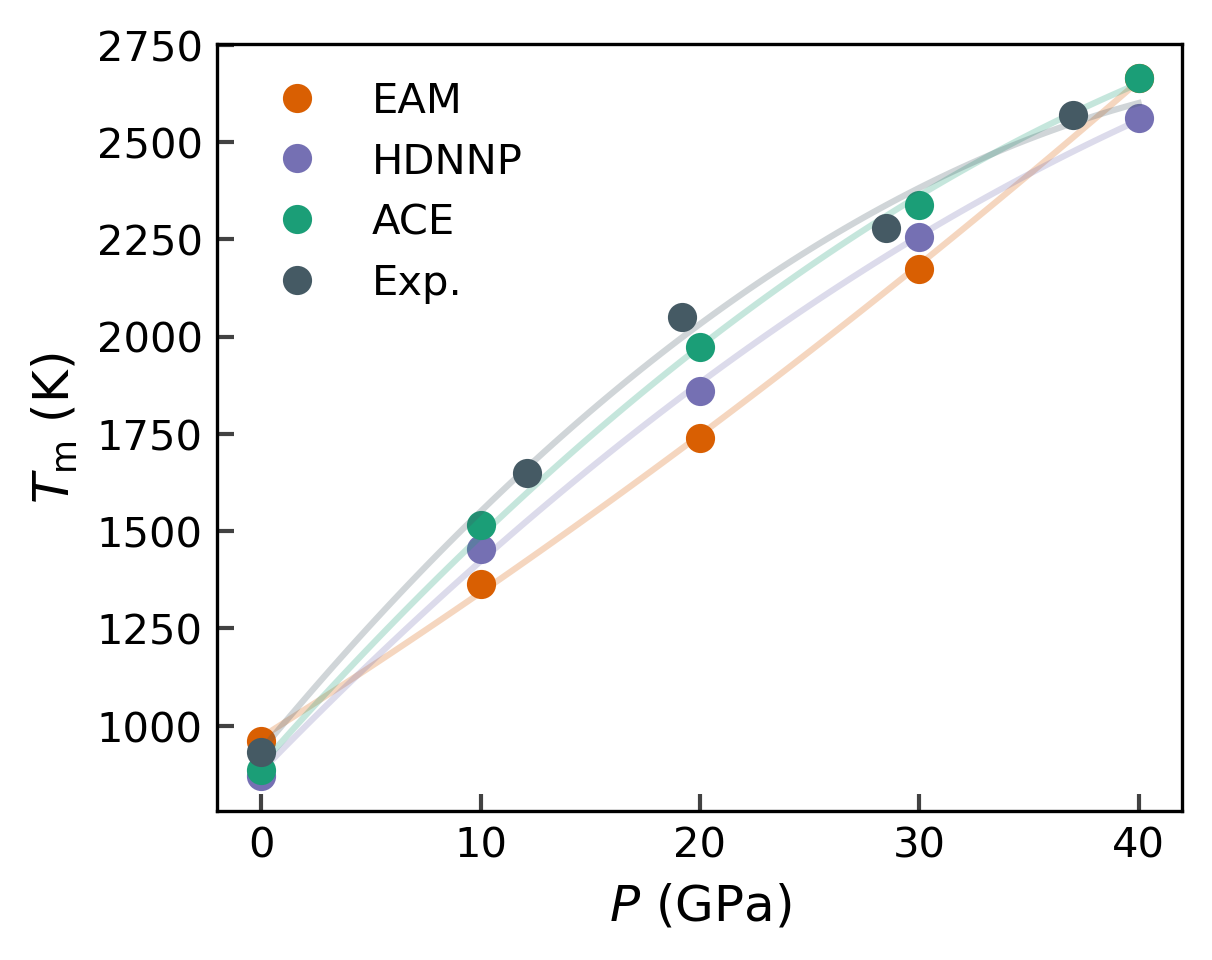}
\caption{Melting curve of fcc Al up to 40 GPa. The melting temperature, $T_m$, at various pressures is calculated for EAM, HDNNP, and ACE. Melting temperatures determined from laser melting experiments \cite{Hanstrom2000} are marked in gray.}
\label{fig:pd_al}
\end{figure}

For pure Al, we present the pressure-temperature $P-T$ phase diagrams.
In order to arrive at the $P-T$ phase diagrams, the free energies of the relevant phases are calculated as a function of temperature and pressure.
To this end, we perform reversible scaling calculations which provide the free energy over a given temperature range.
A pressure range of 0-40 GPa is chosen, with free energy calculations carried out at intervals of 10 GPa.
The fcc and liquid phases are considered, and at each pressure, the melting temperature is obtained from the intersection of the free energy curves. 
A system size of approximately 7000 atoms is chosen for both phases such that any finite size effects are avoided. 
The same set of calculations is performed with all three potentials, the results of which are shown in Fig. \ref{fig:pd_al}.
In general, all three potentials closely follow the predictions from experiments. 
Although the zero pressure melting temperature is underestimated compared to the experimental value \cite{Lide2004}, it is comparable with the melting temperature from \textit{ab initio} calculations \cite{Vocadlo2002}.

For the construction of the binary phase diagram, the fcc and liquid are considered in the composition range $0 \leq x_\mathrm{Li} \leq 0.5$, and B32 AlLi in $0.4 \leq x_\mathrm{Li} \leq 0.5$, and $\mathrm{Al}_3 \mathrm{Li}$ in $0.2 \leq x_\mathrm{Li} \leq 0.3$.
We chose the composition ranges for AlLi and $\mathrm{Al}_3 \mathrm{Li}$ based on the relevant regions in the experimental phase diagram.
We then ascertain that the free energies of these phases were significantly higher than the other phases outside of the selected composition range.
In order to create an Al-rich fcc lattice with Li as impurity, Al atoms are randomly selected and replaced by Li, that is, we assume that Li impurities occupy substitutional lattice positions.
Similarly, substitutional Al impurities are introduced in the B32 structure to create off-stoichiometric compositions.
No other mechanisms, such as vacancies, or interstitials are considered.

Within the selected composition range, we perform free energy calculations at composition intervals of 0.01 for all the phases. 
At each composition, temperatures from 600-1000 K are considered, and the free energy over this range is obtained in a single calculation using the reversible scaling approach. 
Free energy calculations are then performed with timescales of 25 ps for equilibration, 50 ps for switching, and system size of roughly 7000 atoms for each phase.

\begin{figure}[ht]
\centering
\includegraphics[width=0.48\textwidth]{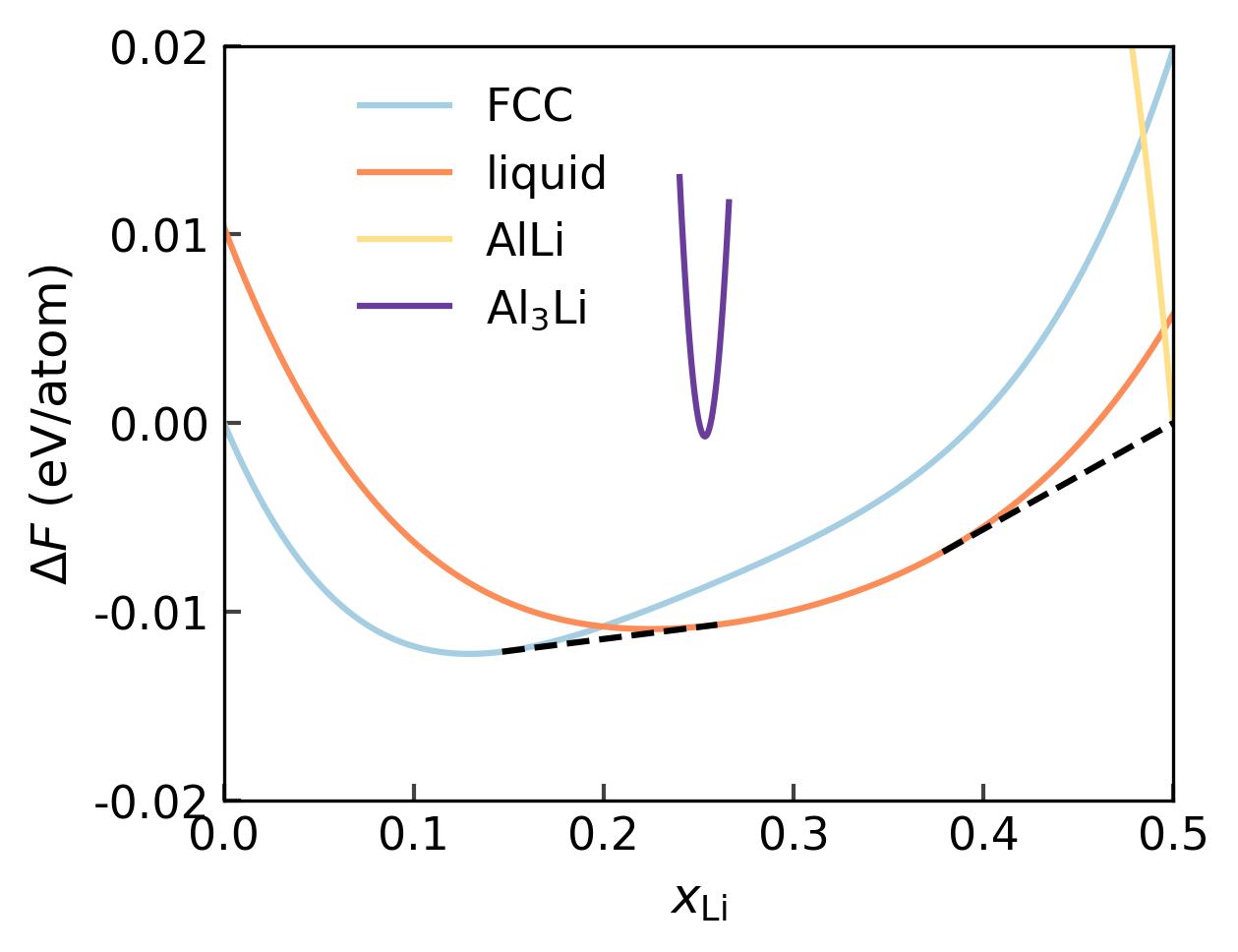}
\caption{Free energy of mixing, $\Delta F$, for fcc, liquid, AlLi, and $\mathrm{Al}_3$Li at 800 K as a function of the composition of Li, calculated with the ACE potential. Substitutional impurity atoms are added in each of the phases to obtain free energy variation with composition (up to 0.5 Li). The two-phase coexisting regions are identified through common tangent constructions, indicated by black dashed lines.}
\label{fig:pd_curves}
\end{figure}

\begin{figure*}[ht]
\centering
\includegraphics[width=0.9\textwidth]{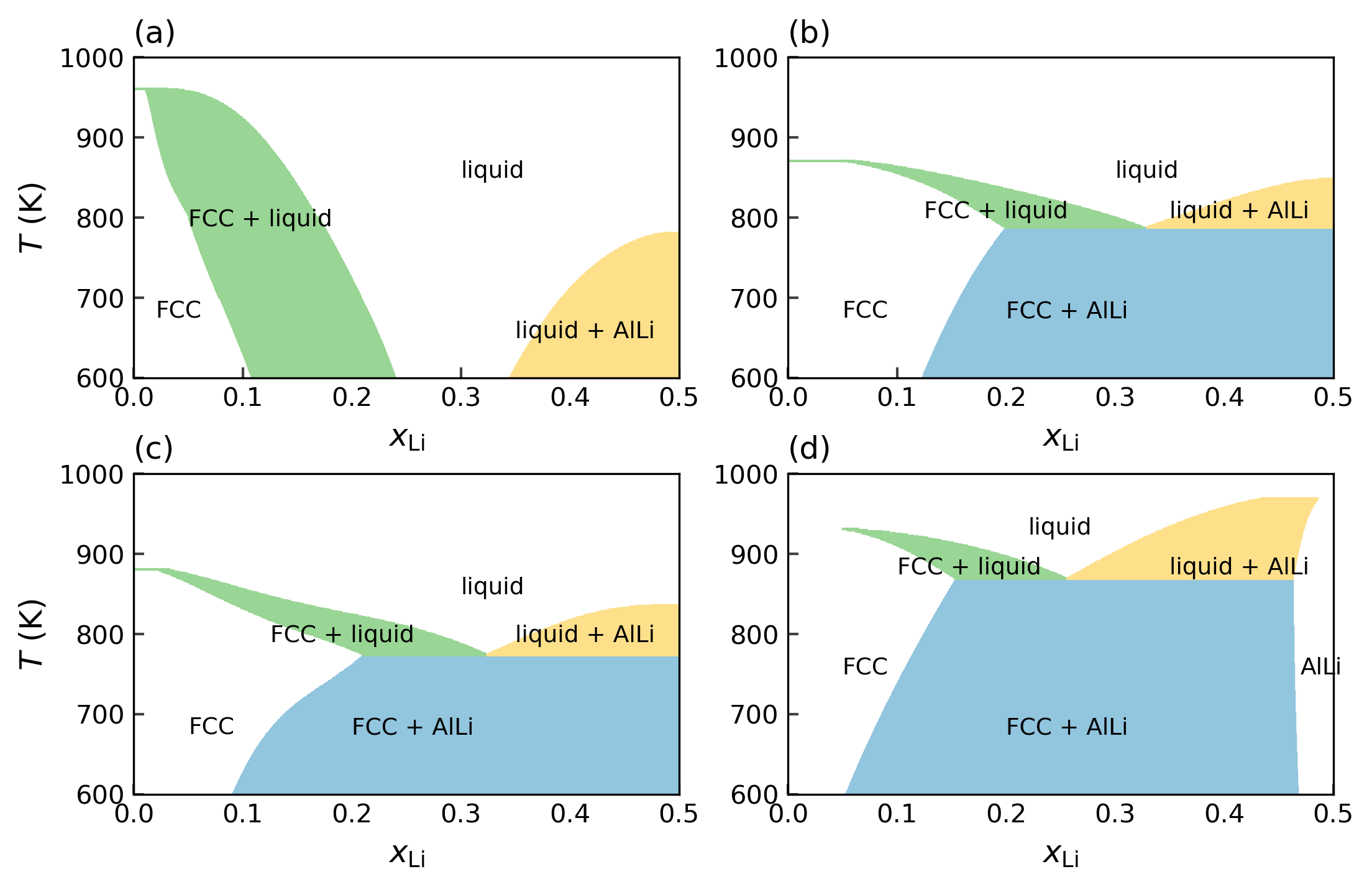}
\caption{Phase diagram of Al-Li up to $x_{\mathrm{Li}}$=0.5, calculated using (a) EAM, (b) HDNNP, and (c) ACE. In (d), the phase diagram calculated using the CALPHAD method is shown.}
\label{fig:pd_complete}
\end{figure*}

Once the free energies are obtained, at each temperature the free energy for each phase with varying composition is extracted, as shown in Fig. \ref{fig:pd_curves}.
A current limitation of our workflow is that it does not include the contribution to the free energy due to configurational entropy in the solid phase, therefore the ideal mixing contributions are added to the fcc, B32,  off-stoichiometric phases.
In order to calculate the free energy of mixing, the end-members are chosen to be the phases with the lowest free energy at $x_\mathrm{Li}=0$ and $x_\mathrm{Li}=0.5$ at the given temperature. 
Finally, the convex hull is calculated at each temperature to extract the regions of stability for each phase. Following such a construction, the regions of phase stability and coexistence can be obtained. Once again, the calculations are performed for all three potentials and the results are shown in Fig. \ref{fig:pd_complete} (a-c). 
The reference phase diagram calculated using the CALPHAD approach as implemented in the \PYC{} \cite{Otis2017} tool, with an AlLi database from Ref. \cite{Wang2011}, is shown in Fig. \ref{fig:pd_complete} (d).

\begin{table*}[]
\centering
\caption{Comparison of the salient features of the phase diagram: the melting temperature of the end members, eutectic temperature, eutectic composition and the Li solubility in fcc Al at the eutectic temperature as predicted by the different interatomic potentials. The melting temperature of fcc Al from DFT calculations is estimated to be 786 - 890 K \cite{Vocadlo2002}, depending on the chosen exchange-correlation functional (see Sec. \ref{sec:dft}). Note that since the eutectic point does not appear in the phase diagram predicted by the EAM potential, the values are an estimation.}
\begin{tabular}{cccccc}
\hline
Potential & Melting temperature & Melting temperature & Eutectic temperature & Eutectic composition & Li solubility \\
 & (K), $x_\mathrm{Li}=0.0$ & (K), $x_\mathrm{Li}=0.5$ & (K) & $x_\mathrm{Li}$ & $x_\mathrm{Li}$\\
\hline
EAM       & 961 & 782 & \textless 500            & 0.25 - 0.35              & 0.1 - 0.2     \\
HDNNP     & 871 & 846 & 786                      & 0.33                    & 0.20         \\
ACE       & 886 & 837 & 771                      & 0.32                    & 0.21         \\
Exp.      & 933 \cite{Lide2004} & 965-991 \cite{Hallstedt2007}  & 871-876 \cite{Hallstedt2007}                  & 0.234 - 0.300 \cite{Hallstedt2007}            & 0.124 - 0.180 \cite{Hallstedt2007}\\
DFT       & 786 - 890 \cite{Vocadlo2002} & & & & \\
\hline
\end{tabular}
\label{tab:phase-dia}
\end{table*}

The HDNNP and ACE exhibit phase diagrams (Fig. \ref{fig:pd_complete} (b, c)) show excellent agreement with both the CALPHAD and the experimental phase diagrams.
The main features of the phase diagram, such as the solubility of Li in the Al lattice, the eutectic point (liquid $\rightarrow$ fcc $+$ AlLi), general shape of the liquidus lines are all well reproduced.
A comparison of the melting temperature of the end members, eutectic temperature, eutectic composition, and the solubility of Li in the fcc Al lattice are shown in Table \ref{tab:phase-dia}.
Both MLPs underestimate the melting temperatures and the eutectic temperature by approximately 15 \%, which is expected from the melting temperature of the underlying DFT reference (\cite{Zhu2017, Zhu2020} and \cite{Dorner2018} with references therein).
The prediction of a lower melting temperature by the MLIPs are also evident at increasing pressure, as seen from Fig. \ref{fig:pd_al}.
Both the eutectic composition and Li solubility, are close to the ranges in experimental observations.

A major difference is the solubility of Al in the AlLi ordered phase. Experimental and CALPHAD show solubility, while HDNNP and ACE predict predictions are on the contrary. This discrepancy could be due to a limitation in the phase diagram calculation workflows rather than the interatomic potentials. In B32 AlLi, experimental studies propose that at lower Li concentrations, the Li atoms exhibit a vacancy mediated diffusion mechanism \cite{Kishio1979}. The exclusion of vacancies in favor of substitutional defects could lead to the low solubility of Al in the AlLi phase.

Although the $\mathrm{Al}_3 \mathrm{Li}$ phase is present on  the DFT convex hull, it does not appear in the calculated phase diagram in the given temperature range. At lower temperatures, the ACE potential predicts regions of coexistence of the FCC and $\mathrm{Al}_3 \mathrm{Li}$, and $\mathrm{Al}_3 \mathrm{Li}$ and AlLi, which disappears around 580 K (See supplemental).

The phase diagram of the EAM potential, as represented in Fig. \ref{fig:pd_complete} (c), does not reproduce the characteristic features of the phase diagram, indicating the possible limitations of an empirical interatomic potential. 
Even though the EAM potential provides the closest estimate of the melting temperature of pure Al as compared to experiments, it predicts the stability of the competing phases incorrectly. Therefore, the applicability of the EAM potential in this study is limited to properties of the pure phases, such as the elastic constants.

Overall, to obtain the phase diagram presented in this work for a potential, we require about 120 molecular dynamics simulations of about 150 ps each, making this approach computationally feasible even with the more expensive MLPs. 
Apart from the selection of relevant phases and temperature ranges, the rest of the workflow can be fully automated, allowing for the calculation of phase diagrams to be a routine task in the lifecycle of interatomic potential development.
We observe that a 1 meV difference in free energy of phases leads to about 20 K difference in the calculated transition temperature;
this, in turn combined with the limitations of DFT in predicting transition temperatures indicate that a 15-20 \% difference in transition temperatures as compared to the experimental phase diagrams is to be expected.
Nevertheless, the calculated phase diagrams are highly beneficial to predict the thermodynamic conditions under which an interatomic potential is reliable, and for the interpretation of the observed phase transformation behavior.

\subsection{Conclusions and outlook}
In conclusion, our presented framework, built upon the \PY{} integrated development environment (IDE), establishes a comprehensive, robust, and user-friendly platform for the development of empirical and machine learning potentials. We have successfully demonstrated its versatility by running all tasks necessary in the development cycle of modern interatomic potentials, covering the creation of systematic Density Functional Theory databases, the fitting of DFT data to various interatomic potentials (EAM, HDNNP, and ACE), and the subsequent validation through a largely automated approach. The power and performance of the framework were exemplified in the computation of a binary composition-temperature phase diagram for the Al-Li alloy system, showcasing its applicability to running highly complex simulation protocols over large datasets consisting of thousands of individual atomic structures and for 
technologically important and complex materials systems.

 The potential applications of the framework presented here are vast. Its user-friendly nature and adaptability make it an accessible and open tool for researchers in diverse fields, offering a streamlined approach to MLP development. Future efforts may focus on expanding the range of potential classes that can be incorporated, further enhancing the flexibility and applicability of the framework to a wide range of materials science challenges requiring complex simulation protocols. Ongoing developments will seek to optimize and automate additional aspects of the MLP development cycle allowing researchers to address even more advanced materials properties needed e.g. to compute defect phase diagrams, thermoelectric behavior, or superconductivity, thus paving the way for more efficient and reproducible research practices. 
 
 We envision our presented framework to act as a foundational platform, inviting researchers to explore and study the opportunities opened by machine learning potentials and their diverse applications. The ongoing commitment to openness, reproducibility, and automation positions our framework as a flexible and expandable basis for innovation and discovery in the quickly expanding landscape of using machine learning approaches in materials science. We therefore encourage the community to actively engage with the provided computer codes and software tools, which are openly provided via github and conda.

\section{Methods}

\subsection{Workflows in materials science}

The last few years showed a tremendous change in how high-performance compute clusters are used: While historically large monolithic codes allowed an up-scaling on an increasing number of cores, with the advent of machine learning a new type of computations becomes more and more important where huge numbers of small and medium-sized jobs running various codes need to be combined to get the final result. A prominent example is the fitting and validation of machine learning potentials described in this paper. The number of DFT calculations needed to get high-quality potentials is in the range of a few ten thousands up to several hundreds of thousands individual DFT calculations. Data management for such a large number of jobs requires not only storing the input and output data but also of their status, i.e., whether they ran successfully, whether they converged, or whether they were aborted. If a jobs fail, they need to be resubmitted and it may be necessary to correct their input. For job sizes of a few 10,000 calculations, even small failure rates make a manual handling inefficient. For these types of advanced calculations automated workflow
systems become almost mandatory. 

In the present work, we have used \PY{} \cite{Janssen2019} as a workflow platform to include all the necessary tools to create advanced machine learning potentials. \PY{} provides features that are well-suited for these tasks: It provides an easy way to run large numbers of DFT calculations (to create the reference data set), to perform the training, and to analyze extensive sets of interatomic potential calculations (for validation). \PY{} provides several features that make running such complex workflows efficient and intuitive for users. Its generic input and output provide an easy way to substitute one DFT code or potential/ML approach with another one. For example, to replace the DFT code the main change would be to change the job type. The generic input specifying the basis set, k-point sampling etc. remains unchanged and will be translated by \PY{} into the code-specific format. The close integration within the Jupyter ecosystem provides interactive and easy access to all workflow components and data, and the availability of advanced job management tools provides an efficient route to upscale and run all calculations on modern supercomputer architectures. 

\subsection{DFT Calculations} \label{sec:dft}

Density Functional Theory (DFT) is a quantum mechanical modeling approach that has become the de facto standard for ab initio computations of materials properties, especially for larger system sizes. This method allows for the calculation of materials properties without the need for fitting or empirical parameters, offering a rigorous and first-principles-based framework for providing the large data sets needed to fit empirical or machine learning potentials. In DFT, a pivotal approximation lies in the exchange-correlation (xc) functional. This approximation enables the reduction of the high-dimensional many-body interaction to a 3D mean field potential incorporated into the Kohn-Sham equations. 

A restriction of all available xc-functionals is that they cannot be systematically improved, i.e., deviations to experiment are inherent. Common functionals, such as the PBE-GGA functional employed in this study, generally demonstrate good agreement with experimental results. However, it is important to note that deviations exist, with errors in bond lengths typically around 1\%, discrepancies in elastic constants potentially reaching 10\% and errors in the melting temperature in the order of 100\,K.\cite{Zhu2017, Zhu2020}

While the exchange-correlation functional represents the only non-controllable approximation in DFT, there exist other parameters that can be used to systematically improve accuracy, albeit at an increased computational cost. Among these, the plane wave energy cutoff, which defines the completeness of the basis set, and the k-point sampling are particularly crucial. Achieving convergence in material properties concerning the choice of these parameters is imperative, especially when employing DFT data for training interatomic potentials. Inadequate convergence does not only lead to often non-systematic deviations from converged results but also introduces noise-like discontinuities in the energy surface due to the discrete nature of the plane wave basis set and the k-point set. For the generation of DFT datasets for potential fitting, it is therefore crucial to carefully select these convergence parameters to ensure that the amplitude of discontinuous fluctuations remains small compared to the targeted error. To be used for development of MLIPs, this typically means an energy convergence to about 1 meV/atom and a force convergence to about 0.1 eV/\AA{}.

When carefully choosing these convergence parameters, DFT is known to smoothly interpolate between similar structures. This characteristic renders DFT particularly well-suited for applications demanding a smooth energy surface and derivatives (e.g. forces and stresses) such as developing interatomic potentials. 

\subsection{Interatomic potentials}

\subsubsection{EAM}

In EAM potentials the energy of the system is given
by a pair potential $V$ and a nonlinear function $F$,
called embedding energy

\begin{equation}
    E = \frac{1}{2} \sum_{ij} V(r_{ij}) + \sum_i F(\rho_i).
\end{equation}

Here, $\rho_i$ is given by $\rho_i = \sum_j \rho(r_{ij})$
and is called electron density.
It is motivated by viewing each atom in a solid as impurity
that is embedded in the host matrix and
therefore subject to its electron density,
leading to attractive chemical interactions.
Then, $V$ can be considered as repulsive core-core interaction. \cite{dawEmbeddedatomMethodDerivation1984}
In modern EAM potentials $V$, $\rho$ and $F$ are chosen
to best reproduce certain properties and do not necessarily
follow the constraints resulting from this motivation,
e.g. $V$ often includes attractive terms.
When freely choosing these function the EAM formalism
is equivalent to the effective-medium \cite{jacobsenInteratomicInteractionsEffectivemedium1987} and Finnis-Sinclaire \cite{finnisSimpleEmpiricalBody1984} potentials.
The potential we fitted closely follows a procedure applied by Mishin et al. \cite{mishinStructuralStabilityLattice2001}.
Details on the employed functional forms and constraints can be found in the
original reference and the supplemental information.

\subsubsection{HDNNP}

The general ansatz underlying the development of second-generation HDNNPs, introduced in 2007 by Behler and Parrinello \cite{Behler2007,Behler2021b}, is that the total energy $E_{\mathrm{tot}}$ of a system can be decomposed into $M$ environment-dependent atomic energy contributions $E_i$, such that
\begin{align}
    E_{\mathrm{tot}} = \sum_{i=1}^{M} E_i(\vec{G_i}(\vec{r}))
    \,.\label{eq:2GEnergyDecomposition}
\end{align}
This approach, which extended the applicability of MLPs to condensed systems containing large numbers of atoms, is based on the assumption that for many systems the atomic energy to a good approximation is a local property that depends only on the interaction of a central atom with its neighboring atoms inside a sphere of radius $r_{\mathrm{c}}$. The environment inside this cutoff sphere is captured by a vector $\vec{G_i}$ of atom-centred symmetry functions, which in turn depend on the coordinates of all neighbours while maintaining the mandatory rotational, translational and permutational invariances. The functional form of these many-body descriptors is described in more detail elsewhere\cite{Behler2011}.
Each entry in $\vec{G_i}$ is passed to an input node of an element-specific dense feed-forward neural network, which provides the atomic energy as its output. \par

During training, the weights of all atomic neural networks are iteratively updated based on the loss gradients of both the total energies and the atomic force components in the training data set to achieve the best match to the reference values in the training set. Further information on the construction and training of HDNNPs can be found in \cite{Behler2015} and \cite{Tokita2023}.

\subsubsection{ACE}

The atomic cluster expansion (ACE) \cite{Drautz2019} introduces basis functions that are complete in the space of atomic environments. In analogy to HDNNPs and other MLPs, the energy is represented by a sum of individual atomic energies within a cutoff sphere, for $N$ atoms,

\begin{equation}
    E_{tot} = \sum_i^{N}E_i .
\end{equation}

The individual energies are calculated from general, abstract atomic properties ($\varphi_i$), which in ACE are expanded as

\begin{equation}
    \varphi_i = \sum_v^{n_v}c_v B_{iv} \,,
\end{equation}

where $c_v$ are the expansion coefficients for the $n_v$ basis functions $B_{iv}$. 
In linear ACE, $E_i$ is written directly as

\begin{equation}
    E_i = \varphi_i \,.
\end{equation}

However, a more efficient approach is to calculate atomic energies as

\begin{equation}
    E_i = \mathcal{F} (\varphi_i^{(1)}, \varphi_i^{(2)}, ..., \varphi_i^{(P)}) \,,
\end{equation}

where $\mathcal{F}$ can be any general non-linear function. The ACE potential used in this work employs a mildly non-linear form with two atomic properties  and a square-root embedding as in the Finnis-Sinclair method

\begin{equation}
    E_i = \varphi_i^{(1)} + \sqrt{\varphi_i^{(2)}} \quad \,.
\end{equation}

\subsection{Thermodynamics} \label{sec:thermo}

One of the most widely employed techniques to calculate free energies through atomistic simulations is thermodynamic integration \cite{Kirkwood1935, Smit2001}.
In this computational technique, a system of interest and a reference system with known free energy are coupled with a parameter $\lambda$.
The Hamiltonian of the combined system is given by

\begin{equation}
    H(\lambda) = \lambda H_f + (1-\lambda) H_i
\end{equation}

where $H_i$, is the initial or reference system with the known free energy, and $H_f$ is the final system, or the system of the interest.
If the system of interest is in the solid state, we use a non-interacting Einstein crystal \cite{Frenkel1984} as the reference state, while for liquids, an Uhlenbeck-Ford model \cite{PaulaLeite2016} is employed.
The free energy difference between the two systems can be calculated as

\begin{equation}
    F_f = F_i = \int_{\lambda=0}^{\lambda=1} d\lambda \bigg \langle \frac{\partial H(\lambda)}{\partial \lambda} \bigg \rangle_\lambda .
\end{equation}

The integration has to be performed over a discretized $\lambda$ array, and therefore is computationally quite expensive, which calls for methodological improvements.
In the non-equilibrium approach to thermodynamic integration \cite{Watanabe1990}, the coupling parameter $\lambda$ is time-dependent, and the switching between the initial and final system is carried out in both forward and reverse directions in a single time-dependent calculation. 
The work done in such a switching process is calculated as,

\begin{equation}
  W^{s} = \int_{t_i}^{t_f}
  \frac{d \lambda(t)}{dt}
  \frac{\partial H (\lambda)}{\partial \lambda}  dt 
\end{equation}

which is related to the free energy difference $\Delta F$ between the two systems, 

\begin{equation} \label{eq:df2}
  \Delta F = W^{rev} = {{W}^{s}} - {{E}^{d}}  .
\end{equation}

$E^d$ is the energy dissipation in the switching process, which can be obtained as the difference between the forward and reverse switching.
The non-equilibrium approach can be used to efficiently calculate the free energy of the system of interest at a given thermodynamic condition ($P$, $T$). Once a free energy $F(P, T)$ is known the free energy as a function of temperature over a given range from $T$ to $T_f$ can be obtained in a single calculation using the reversible scaling approach \cite{deKoning1999}. These approaches, and associated algorithms have been discussed in more detail in Ref. \cite{Menon2021}.

\subsection{Software}

\subsubsection{\PY{}}

\PY{} is a workflow framework for atomistic simulation, focused on rapid prototyping and up-scaling simulation protocols. Based on an object-oriented approach, the individual components of a simulation protocol in \PY{} are combined like building blocks. Each \PY{} object is connected to the jupyter-based user-interface, the data storage interface which combines a structured database (SQL) and a hierarchical file format (HDF5) as well as the resource interface to connect to computing resources and parameter databases. By implementing the potential fitting codes (\texttt{atomicrex}, \texttt{RuNNer} and \texttt{pacemaker}), the simulation codes (\texttt{LAMMPS} and \texttt{VASP}) and the thermodynamics code (\texttt{calphy}) based on the same job class, the technical complexity of executing the underlying codes is greatly reduced.

As a first step of the simulation protocol a new project is initialized: \mintinline{python}{pr = Project("AlLi")}. The project object is represented as a folder on the file system and all calculations in this project are going to be executed in this folder. From the project object the individual job objects are created using the factoring pattern:

\begin{minted}{python}
job = pr.create.job.SimulationCode('job_name')
\end{minted}

The factoring pattern, which refers to using one object to create objects of different types, has two advantages: On the one hand it allows the users to use auto-completion in selecting the new object to create and on the other hand the newly created object can be already initialized with information of the object it is created from. In this case the job object receives its storage location from the project object is was created from. The individual job classes for the \texttt{VASP} DFT code, the different fitting codes and \texttt{calphy} and \texttt{LAMMPS} for validation are introduced below. 

\subsubsection{\texttt{PyXtal}} \label{sec:pyxtal_workflow}

For the generation of random crystal structures we have wrapped the python code \texttt{PyXtal} in the \texttt{structuretoolkit} module distributed with pyiron.

\begin{minted}{python}
import structuretoolkit.build.random as stkr
al_li_structures = stkr.pyxtal(
    group=[227, 194], 
    species=["Al", "Li"], 
    num_ions=[4, 4], 
    repeat=10
)
\end{minted}

would generate a list of ten structures each of the spacegroups 227 and 194 with stoichoimetry Al$_4$Li$_4$.
More advanced options as document by the \texttt{PyXtal} library itself, can be passed to the function as well.

\subsubsection{\texttt{VASP}} \label{sec:vasp_workflow}
Starting with the Vienna Ab initio Simulation Package (VASP) \cite{kresse1993ab,kresse1996efficiency,kresse1996efficient}, the job object is created from the project object using the factoring pattern and an atomic structure in the \mintinline{python}{Atoms} format defined by the Atomic Simulation Environment (ASE) is assigned: 
\begin{minted}{python}
job = pr.create.job.Vasp("job_name")
job.structure = structure
\end{minted}
In addition to the atomic structure also the input parameters which determine the precision of the DFT calculation can be specified directly through the \PY{} python interface. For this \PY{} provides two interfaces, first the generic interface which is independent of the specific simulation code and second the code-specific interface, which allows users already experienced with a specific simulation code to directly modify specific input parameters. Using the generic interface the plane wave energy cutoff is set to $750$ eV, the k-point density is set to $0.1$ \AA{}$^{-1}$ and the level of electronic convergence is defined as $10^{-8}$ eV: 
\begin{minted}{python}
job.set_encut(750.0)
job.set_kpoints(k_mesh_spacing=0.1)
job.set_convergence_precision(
    electronic_energy=1.0e-8,
)
job.set_occupancy_smearing(
    smearing="FermiDirac", 
    width=0.2,
)
\end{minted}

The advantage of using the generic interface is that the users can switch between different DFT simulation codes by only changing the create job function call \mintinline{python}{pr.create.job.Vasp()}, the rest of the commands remain the same. For expert users \PY{} also provides the option to access the simulation code specific input directly. As an example, while the electronic smearing can be specified using the generic \mintinline{python}{set_occupancy_smearing()} function, it can also be modified based on the \texttt{VASP} specific input file named \texttt{INCAR}, which can be accessed in \PY{} like a python dictionary: 
\begin{minted}{python}
job.input.incar["ISMEAR"] = -1
job.input.incar["SIGMA"] = 0.1
\end{minted}
Finally, in addition to the simulation code-specific parameters the \PY{} job object also provides the option to specify the submission to the high performance computing (HPC) queuing system:
\begin{minted}{python}
job.server.queue = "gpu_queue"
job.server.cores = 4
job.server.gpus = 4
\end{minted}
After the specification of the input parameters and resource assignment is completed the \PY{} job object can be executed using the \mintinline{python}{run()} function. This triggers the internal cycle of writing the input files, submitting the calculation to the HPC for execution and once the calculation is completed parse the output files to provide the output to the pyiron python interface. 

\subsubsection{TrainingContainer} \label{sec:tc}

Following the execution of the DFT calculations, the next step is the aggregation of the outputs of these calculation to provide them to the fitting codes for the interatomic potentials. In \PY{} this is achieved by combining two objects, the \PY{} \mintinline{python}{table} object and the \mintinline{python}{TrainingContainer}. The \PY{} \mintinline{python}{table} object specifies a series of functions which are applied to each job object in a given \PY{} project, following a map-reduce pattern: 
\begin{minted}{python}
table = pr.create.table()
table.add.get_job_name
table.add.get_structure
table.add.get_energy_tot
table.add.get_forces
table.run()
\end{minted}
The aggregated data, which is returned as a pandas DataFrame object is then stored in the \mintinline{python}{TrainingContainer} for reference in the fitting codes: 
\begin{minted}{python}
tr = pr.create.job.TrainingContainer("tc_job")
tr.include_dataset(table.get_dataframe())
\end{minted}

Additionally, the class defines common plotting that make the creation of graphs such as Figs. \ref{fig:convexhull} and \ref{fig:eV_curves} easier, for example,

\begin{minted}{python}
tr.plot.energy_volume()
\end{minted}

\subsubsection{\AT{}}

The \AT{} interface exposes the full functionality of the code \cite{stukowskiAtomicrexGeneralPurpose2017} in a \PY{} python interface,
while storing relevant inputs and output necessary to reproduce fitting processes.
An \texttt{atomicrex} job object can be created with
\begin{minted}{python}
job = pr.create.job.Atomicrex("AtomicrexJob")
\end{minted}
Currently \AT{} implements EAM, modified EAM \cite{baskesApplicationEmbeddedAtomMethod1987},
angular dependent \cite{mishinPhaseStabilityFe2005},
analytic bond order \cite{brennerEmpiricalPotentialHydrocarbons1990} and Tersoff \cite{tersoffEmpiricalInteratomicPotential1988} potentials.
They can be set using
\begin{minted}{python}
pot = job.factories.potentials.potential_type()
\end{minted}
For potentials that allow for different functional forms like EAM potentials
it is necessary to define these functions.
Here the user can choose between predefined functions and own creations via a math parser.
\begin{minted}{python}
morse = job.factories.functions.morse_B(
    identifier="V", 
    D0=0.05, 
    r0=2.5, 
    beta=2.2, 
    S=2.4, 
    delta=0.0, 
    species=["Al", "Li"]
)
uf = ref.factories.functions.user_function(
    identifier="UserElement1Element2", 
    input_variable="r"
)
uf.expression = "A*exp(r0-r)"
uf.derivative = "-A*exp(r0-r)"
uf.parameters.add_parameter("A", 3)
uf.parameters.add_parameter("r0", 5)
pot.pair_interactions[morse.identifier] = morse
pot.electron_densities[uf.identifier] = uf
\end{minted}
Structures and corresponding fit properties can be directly assigned using the general \texttt{TrainingContainer} interface.
If fine grained control over weights is required they can also be added one by one:
\begin{minted}{python}
s = pr.create.structure.ase.bulk("Al")
job.structures.add_structure(s, 
    identifier="SomeStructure", 
    relative_weight=10000)
job.structures.add_scalar_fit_property(
    "atomic-energy",
    target_val=-4.0,
    relative_weight=100,
)
\end{minted}
Nearly arbitrary parameter constraints can be added using math parser expressions:
\begin{minted}{python}
job.input.parameter_constraint.add_constraint(
    identifier="SomeContstraint",
    dependent_dof="constrainedParameter",
    expression="MathparserExpression",
)
\end{minted}
Finally, the user can choose between an internal LBFGS minimizer
and a plethora of optimization algorithms provided via the NLopt library \cite{NLopt} to fit the potential.
\begin{minted}{python}
algo = job.factories.algorithms.some_algo(
    max_iter=1000
)
job.input.fit_algorithm = algo
\end{minted}

\subsubsection{\RU{}}

Training with \RU{} usually passes through three stages: in mode 1, the values of the atom-centered symmetry functions for the whole training dataset are calculated and stored to disk, and the data is separated into a training and a test set. mode 2 optimizes the parameters of the HDNNP in order to represent best the reference energies and forces. Finally, mode 3 is used to predict the properties of unknown configurations. 

The \PY{} job {\texttt{\detokenize{RuNNerFit}}} reflects these steps. Similar to the other training jobs, it is created by invoking the {\texttt{\detokenize{create}}} routine of a \PY{} {\texttt{\detokenize{Project}}} object. Every {\texttt{\detokenize{RuNNerFit}}} job also requires the specification of a training dataset.

\begin{minted}{python}
mode1 = proj.create.job.RunnerFit('mode1')
mode1.add_training_data(dataset)
\end{minted}

In the next step, a set of atom-centered symmetry functions must be parameterized for the training dataset. {\texttt{\detokenize{runnerase}}} offers the procedure {\texttt{\detokenize{generate_symmetryfunctions}}} to help with this task. Afterwards, the job can be started using the {\texttt{\detokenize{run}}} command:

\begin{minted}{python}
sfs = generate_symmetryfunctions(dataset, 
    sftype=2, 
    cutoff=12.0
)
mode1.parameters.symfunction_short += sfs
mode1.run()
\end{minted}

After the successful termination of mode 1, mode 2 is started by reloading the first job and altering the setting {\texttt{\detokenize{runner_mode}}}. This tells the underlying \texttt{RuNNer} code how to operate:

\begin{minted}{python}
mode2 = mode1.restart('mode2')
mode2.parameters.runner_mode = 2
mode2.run()
\end{minted}

The same procedure is followed to run mode 3. In a {\texttt{\detokenize{RuNNerFit}}}, the execution of mode 3 is mandatory to complete training and obtain a full prediction of both the train and test datasets:

\begin{minted}{python}
mode3 = mode2.restart('mode3')
mode3.parameters.runner_mode = 3
mode3.run()
\end{minted}

In order to use the trained potential in an application with \texttt{LAMMPS}, one can call the {\texttt{\detokenize{get_lammps_potential}}} routine which returns the required {\texttt{\detokenize{pair_style}}} and {\texttt{\detokenize{pair_coeff}}} commands. The HDNNP pair style is part of the \texttt{LAMMPS} interface provided by the \texttt{n2p2} package \cite{n2p2}:

\begin{minted}{python}
mode3.get_lammps_potential()
\end{minted}

\subsubsection{\PM{}}

In order to setup the \PM{} job, one needs to create the corresponding \PY{} job and add the training dataset.
\begin{minted}{python}
job = pr.create.job.PacemakerJob("pacemaker_job")
job.add_training_data(dataset)
\end{minted}
Parameters for ACE parameterization procedure will be initialized to their defaults. However, one always can configure all of them. For example, setting energy-force weights balance ($\kappa$ from~Ref.\cite{Bochkarev2022}) as 
\begin{minted}{python}
job.input['fit']['loss']['kappa']=0.3
\end{minted}
After that one can run the job and get the \texttt{LAMMPS} potential as well.

\subsubsection{\CA{}} \label{sec:calphy}

The computational approaches to obtain free energies as discussed in Section \ref{sec:thermo} consists of multiple interdependent steps, and presents a complex computational workflow.
In order to facilitate a user to easily calculate the free energies, and at the same time retain the ability to tune each step in the workflow as needed, we developed \CA{} \cite{Menon2021}, a python library for automated calculation of free energies.
It uses \LA{} as the molecular dynamics driver to perform free energy calculations in an automated manner.
\CA{} when combined with \PY{}, can leverage additional features such as interoperability with other common atomistic simulation tools, scaling to HPC systems, and job and data management.

Within \PY{}, a non-equilibrium free calculation, for example an Al fcc lattice at 500 K and 0 pressure can be carried out by the following code:

\begin{minted}{python} 
pr = Project("free_energy")
job = pr.create.job.Calphy("Al_fcc_500")
job.structure = pr.create.structure.ase.bulk("Al", 
    cubic=True).repeat(4)
job.potential = "Al-atomicrex"
job.calc_free_energy(
    temperature=500,
    pressure=0,
    reference_phase="solid",
    n_equilibration_steps=25000,
    n_switching_steps=50000,
)
job.run()
\end{minted}

The main inputs needed are the input structure and the interatomic potential, apart from the thermodynamic conditions at which the calculation is to be performed.
For calculating the free energy of a liquid system, the only change needed is \mintinline{python}{reference_phase='liquid'}. \CA{} automatically uses a different reference system based on this command. 
To obtain free energies over a given temperature range,  one needs to change the temperature option: \mintinline{python}{temperature=[500, 800]}. 
In this case, a free energy calculation at 500 K is performed first, followed by a temperature integration up to 800 K in another calculation.

\subsubsection{\LA{}}

Beyond the free energies calculated with \CA{} to construct the phase diagram, the \texttt{LAMMPS} molecular dynamics simulation code is used to validate material properties calculated with the individual machine learning potentials. In pyiron the workflows to calculate the material properties is defined independent of the specific simulation code, so in the first step a reference \texttt{LAMMPS} job is defined for the interatomic potential fitted with the \texttt{atomicrex} fitting code: 
\begin{minted}{python}
pr = Project('validation')
job = pr.create.job.Lammps('lmp')
job.structure = structure
job_lmp.potential = 'Al-atomicrex'
\end{minted}
Following the definition of the reference job the next step is assigning this reference job to the workflow to calculate a material property, in this case the calculation of the elastic constants with the \mintinline{python}{ElasticMatrix} job:
\begin{minted}{python}
elastic = pr.create.job.ElasticMatrix('elmat')
elastic.ref_job = job_lmp
elastic.run()
\end{minted}
By defining the calculation of the material properties independent of the simulation code, the same validation calculation can be applied for the \texttt{LAMMPS} molecular dynamics simulation code to test the fitted interatomic potentials as well as the \texttt{VASP} DFT simulation code, to enable a direct comparison.  

\subsection{Software and data availability}

The software used in this paper, \PY, \PM, \RU, \AT, \CA, \LA, \PYC, and \texttt{PyXtal}  
~are freely available from their respective repositories. 
A list of the software tools, along with their repositories and documentation is provided in the supplementary material.
Exemplary workflows to illustrate the calculations mentioned in this manuscript are available in an online repository\cite{WorkflowRepo}, along with the free energy values for the construction of the phase diagram.
In addition, the dataset used for parametrization of the interatomic potentials, is also made available \cite{DatasetTemporary}.

\subsection{Acknowledgements}

The workflows, potentials, and results presented here were obtained in the framework of the POTENTIALS collaboration and scientific network ``Assessment of atomistic simulations'' with funding from the German Science Foundation (DFG) (grant number 405602047). 
Furthermore, a workshop on the subject of this manuscript at which participants could interactively execute and explore the initial versions of these workflows was held in June 2022 \cite{PotentialWorkshop}.

S. M. acknowledges funding by the Deutsche Forschungsgemeinschaft (DFG, German Research Foundation) under the National
Research Data Infrastructure – NFDI 38/1 – project number 460247524. J. B. acknowledges funding by the DFG (project number 405479457 as part of PAK 965/1). A. K. acknowledges funding by the Studienstiftung des Deutschen Volkes (doctoral scholarship). N. L. and J. R. acknowledge funding by the Deutsche Forschungsgemeinschaft (DFG, German Research Foundation) under grant number 405621137. K. A. acknowledges funding from the the DFG under grant number 405621160. M. M. and R. D. acknowledge funding by the German Science Foundation (DFG), projects 405621081 and 405621217. R.D. and Y. L. acknowledge computation time by Center for Interface-Dominated High Performance Materials (ZGH) at Ruhr-Universität Bochum, Germany. J.J. and J.N.  acknowledge funding by the DFG under grant number 405621217. M.P. and J.N.  acknowledge funding from the DFG under grant number 405621160.

\bibliography{references} 

\end{document}


\title{Supplemental material for \lq From electrons to phase diagrams with classical and machine learning potentials: automated workflows for materials science with \texttt{pyiron}\rq}

\author{Sarath Menon \orcidlink{0000-0002-6776-1213}} 
\email[]{s.menon@mpie.de}
\affiliation{Max-Planck-Institut f\"ur Eisenforschung GmbH, 40237 Düsseldorf, Germany}

\author{Yury Lysogorskiy }
\affiliation{ICAMS, Ruhr-Universit\"at Bochum, 44801 Bochum, Germany}

\author{Alexander L. M. Knoll}
\affiliation{Lehrstuhl f\"ur Theoretische Chemie II, Ruhr-Universität Bochum, 44780 Bochum, Germany}
\affiliation{Research Center Chemical Sciences and Sustainability, Research Alliance Ruhr, 44780 Bochum, Germany}

\author{Niklas Leimeroth \orcidlink{0009-0005-3906-4751}}
\affiliation{Technische Universit\"at Darmstadt, Fachbereich Material und Geowissenschaften, Fachgebiet Materialmodellierung, 64287 Darmstadt, Germany}

\author{Marvin Poul \orcidlink{0000-0002-6029-8748}}
\affiliation{Max-Planck-Institut f\"ur Eisenforschung GmbH, 40237 Düsseldorf, Germany}

\author{Minaam Qamar \orcidlink{0000-0002-3342-4307}}
\affiliation{ICAMS, Ruhr-Universit\"at Bochum, 44801 Bochum, Germany}

\author{Jan Janssen \orcidlink{0000-0001-9948-7119}}
\affiliation{Max-Planck-Institut f\"ur Eisenforschung GmbH, 40237 Düsseldorf, Germany}

\author{Matous Mrovec \orcidlink{0000-0001-8216-2254}}
\affiliation{ICAMS, Ruhr-Universit\"at Bochum, 44801 Bochum, Germany}

\author{Jochen Rohrer \orcidlink{0000-0002-4492-3371}}
\affiliation{Technische Universit\"at Darmstadt, Fachbereich Material und Geowissenschaften, Fachgebiet Materialmodellierung, 64287 Darmstadt, Germany}

\author{Karsten Albe \orcidlink{0000-0003-4669-8056}}
\affiliation{Technische Universit\"at Darmstadt, Fachbereich Material und Geowissenschaften, Fachgebiet Materialmodellierung, 64287 Darmstadt, Germany}

\author{Jörg Behler \orcidlink{0000-0002-1220-1542}}
\affiliation{Lehrstuhl f\"ur Theoretische Chemie II, Ruhr-Universität Bochum, 44780 Bochum, Germany}
\affiliation{Research Center Chemical Sciences and Sustainability, Research Alliance Ruhr, 44780 Bochum, Germany}

\author{Ralf Drautz \orcidlink{0000-0001-7101-8804}}
\affiliation{ICAMS, Ruhr-Universit\"at Bochum, 44801 Bochum, Germany}

\author{J\"org Neugebauer \orcidlink{0000-0002-7903-2472}}
\email[]{neugebauer@mpie.de}
\affiliation{Max-Planck-Institut f\"ur Eisenforschung GmbH, 40237 Düsseldorf, Germany}

\date{\today}

\maketitle

\section{Data Set Generation}

\subsection{Domain-driven data set generation}
First generation of training dataset contained following structures:
\begin{enumerate}
\item Perfect unary crystals (fcc/bcc/hcp) of Al and Li. For each structure, the following properties were computed: an
    energy-nearest neighbor distance curve (from 2 to 6.5\,\AA{} with a 0.5\,\AA{} step), full structural relaxation, an
    energy-volume curve around equilibrium volume ($\pm$10\,\% with a 2\,\% step), elastic matrix calculations (with 5
    points along each deformation mode in a $\pm$0.5\,\% strain range), phonons (as determined by Phonopy \cite{phonopy-phono3py-JPCM, phonopy-phono3py-JPSJ}), and
    a supercell with a single vacancy.

\item Binary prototypes from the Materials Project, that contains Al and Li: Li2Al mp-1210753, LiAl mp-1067, LiAl3 mp-10890, Li9Al4 mp-568404, LiAl mp-1079240, LiAl mp-1191737, Li3Al2 mp-16506). For each structure, the same steps as in p.1 were performed. 

\item Randomly deformed supercells. For each of the optimized structures from pp. 1 and 2, a new supercell was
    constructed in such a way that its minimal length of the cell vector was more than 7\,\AA. For each supercell, five
    random deformations were generated. Each deformation consists of random normal atom displacements with
    $\sigma=0.05$\,\AA{} and a random normal cell deformation with $\sigma=0.05$. For each of the five random
    deformations, 11 uniformly isotropic deformations from -10\,\% to +10\,\% with a 2\,\% step were generated.



\end{enumerate}

This dataset was utilized to train the zeroth generation of the ACE potential. 
An active learning procedure was employed with this potential to sample more configurations. 
New configurations were generated through MD simulations in the NPT ensemble with zero pressure and increasing temperatures from 1 to 1500\,K over 15,000 steps of supercells from p.3.
Only structures with a maximum per-atom extrapolation grade exceeding 5 were selected. 
If the extrapolation grade exceeded 20, simulations were halted.
Extrapolation grades were computed every 5th MD step. 
The number of captured configurations for different crystal structure types ranged from 6 to 261. 
In total, 491 structures were collected during the first round of active learning.
These structures were computed with DFT, added to the training set, and the ACE potential was retrained. 
In the second round of active learning, the same procedure as before was applied, but MD ran for 50,000 steps with a steady temperature increase from 1 to 1500\,K, followed by an additional 50,000 steps at T=1500\,K. A total of 225 configurations were collected, ranging from 4 to 146 configurations per crystal structure.


\subsection{Random Crystal Structures}

Table~\ref{tab:spg_parameter} show the parameters used to generate the training data.
During all steps of this procedure we remove structures that have atomic
distances below $1.9\,\mathrm{\AA}$ to avoid overlapping PAW spheres, which
would negatively impact the quality of the training data.

\begin{table}
    \caption{\label{tab:spg_parameter}\ 
        Hyperparameters for systematic training set generation.
        Within each step of the random perturbations all listed modifications are applied together.
        The number of of samples per structures refers to the number of
        displaced, strained or sheared structures per structure obtained from
        PyXtal after minimization.
    }
    \begin{tabular}{cc|cc}
        Step   & parameter & \multicolumn{2}{c}{value} \\
        \midrule
        PyXtal & \#Atoms, unary  &           & total $\leq 10$ \\
               & \#Atoms, binary & 1,2,3,4,6 & total $\leq 8$ \\
               & \#Samples       & \multicolumn{2}{c}{3346} \\
        \midrule
        Vibrational & displacements & gaussian    & $\sigma=0.2$\,\AA{} \\
                    & strain        & uniform     & 0.05 \\
                    & shear         & uniform     & 0.005 \\
                    & \#Samples     & \multicolumn{2}{c}{4 per structure} \\
        \midrule
        Elastic, Tri-axial & train     & uniform     & 0.05 \\
                           & shear     & uniform     & 0.005 \\
                           & \#Samples & \multicolumn{2}{c}{4 per structure}\\
        \midrule
        Elastic, Shear & strain        & uniform     & 0.05 \\
                       & shear         & uniform     & 0.005 \\
                       & \#Samples     & \multicolumn{2}{c}{4 per structure} \\
    \end{tabular}
\end{table}

\begin{figure}
    \centering
    \includegraphics[width=\columnwidth]{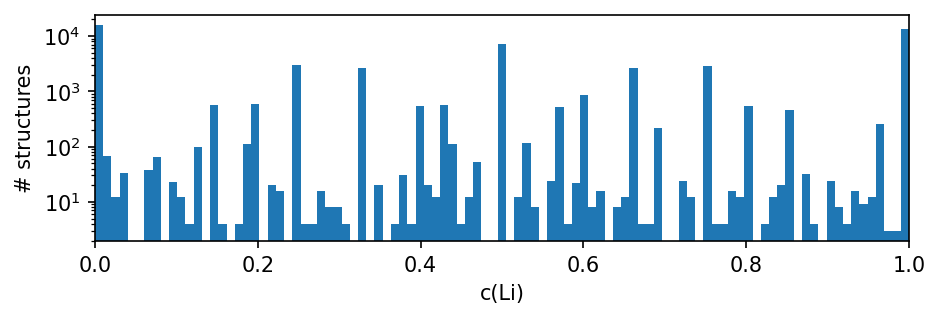}
    \caption{\
        Histogram of Li concentration in the training set.
        y-axis gives the number of structures on a log scale.
    }
    \label{fig:training_concentration}
\end{figure}

\begin{figure*}
    \centering
    \includegraphics[width=\textwidth]{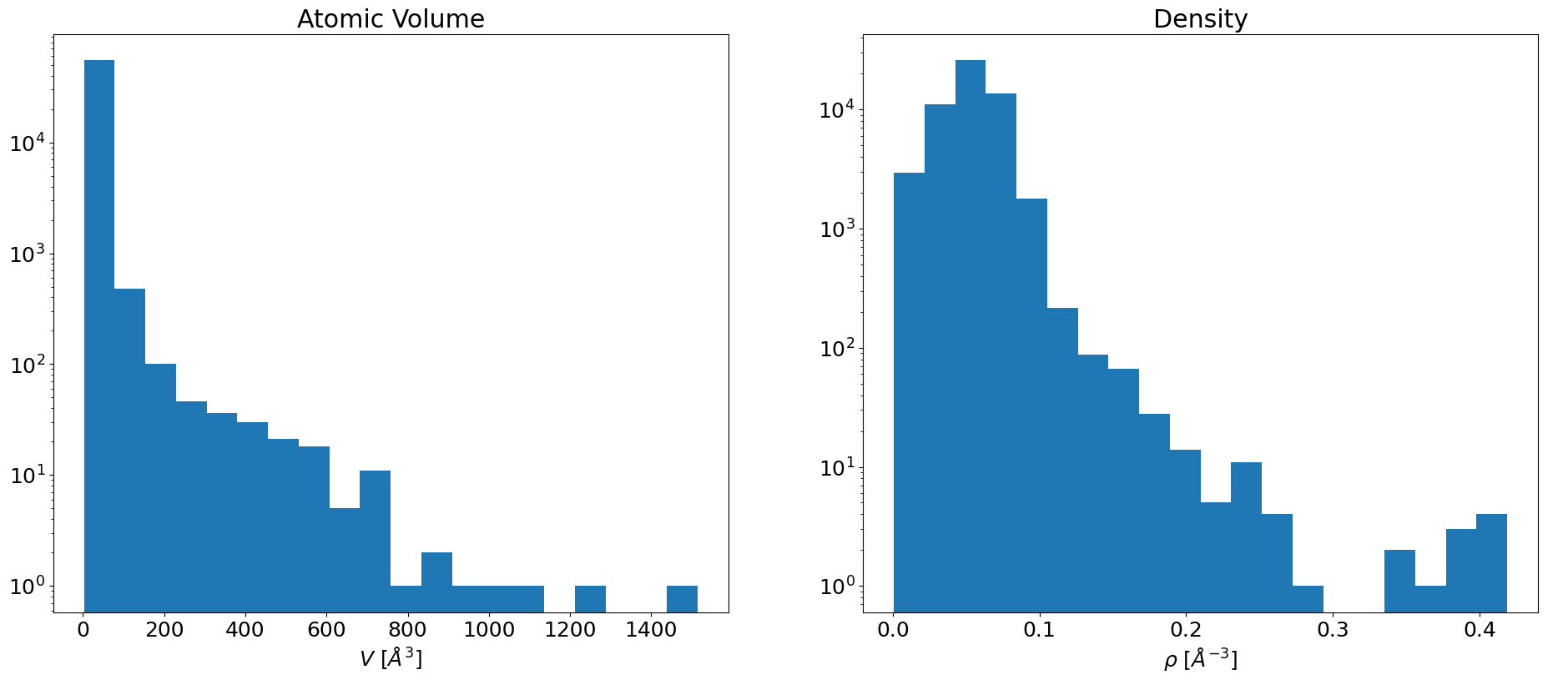}
    \caption{\
        Distribution of (per atom) volume and density in the training set.
    }
    \label{fig:training_size}
\end{figure*}

\begin{figure}
    \centering
     \includegraphics[width=0.4\textwidth]{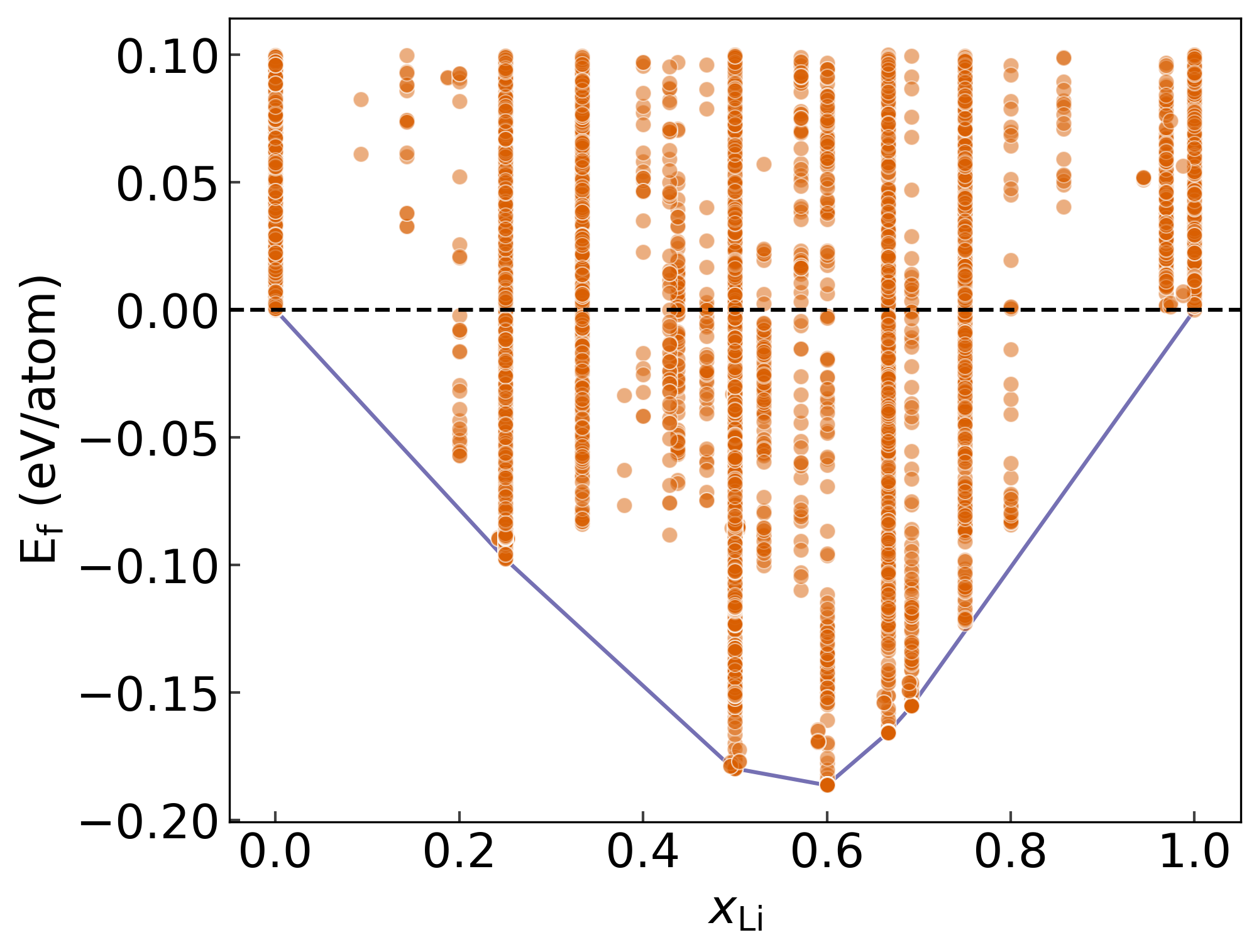}
    \caption{Formation energies ($E_\mathrm{f}$) of atomic configurations included in the training dataset (only configurations with energies below 0.1\,eV are shown for clarity).}
    \label{fig:train_convex_hull}%
\end{figure}

Figure~\ref{fig:training_concentration} shows the distribution of Li
concentration in the \emph{full} training set, Fig.~\ref{fig:training_size} the distribution of volume and Fig.~\ref{fig:train_convex_hull} the convex hull.

\section{Functions in EAM potential}

The pair function $V$ is defined as sum of 2 Morse functions $M(r, r_0,\alpha)=\exp(-2\alpha(r-r_0))-2\exp(-\alpha (r-r_0)$, additional
short range repulsive terms $R(r_s, S) = S(r_s-r)^4~\mathrm{for} ~ r<r_s~\mathrm{else}~0$
and a cutoff function $\Psi(x)=x^4/(1+x^4) \mathrm{for} ~ x\geq 0~\mathrm{else}~0$:
\begin{multline}
     V(r) = \left[ E^{(1)}M(r,r_0^{(1)}, \alpha^{(1)}) + E^{(2)}M(r,r_0^{(2)}, \alpha^{(2)}) \right] \\
     \Psi\left(\frac{r-r_c}{h}\right)+\sum_{n=1}^3 R(r, r_s^{(n)}, S^{(n)}).
\end{multline}
The electron density $\rho$ is given by
\begin{multline}
    \rho(r) = \left[\alpha \exp(-\beta^{(1)} (r-r_0^{(3)})^2) + \exp(-\beta^{(2)}(r-r_0^{(4)})) \right] \\
    \Psi\left(\frac{r-r_c}{h}\right),
\end{multline}
where $\alpha$ is a prefactor used to normalize the total electron density on an atom in the equilibrium structure to 1
with $\overline{\rho} = \sum_m N_m \rho_m=1$,
where $N_m$ is the number of atoms at distance $r_m$ and
$\rho_m$ is $\rho(r_m)$.
The embedding term is defined separately for $\overline{\rho}<1$
\begin{equation}
    F(\overline{\rho})=F^{(0)}+\frac{1}{2}F^{(2)}(\overline{\rho}-1)^2+\sum_{n=1}^4q^{(n)}(\overline{\rho}-1)^{n+2}
\end{equation}
and $\overline{\rho}>1$
\begin{equation}
    F(\overline{\rho})=\frac{F^{(0)}+\frac{F^{(2)}}{2}(\overline{\rho}-1)^2+q^{(1)}(\overline{\rho}-1)^3 + Q^{(1)}(\overline{\rho}-1)^4}{1 + Q^{(2)}(\overline{\rho}-1)^3}.
\end{equation}
To satisfy an exact match of the lattice constant $V$ was constrained by solving $\sum_m N_m R_m V'_m = 0$ for $E^{(1)}$.
An exact match of the cohesive energy $E_0$ is achieved by setting $F^{(0)}=E_0-1/2 \sum_m N_m V_m$.
The expression for the bulk modulus $B$ is obtained from
\begin{equation}
    \frac{1}{2} \sum_m N_m V''_m R^2_m + F^{(2)} \left(\sum_m N_m \rho'_m R_m \right)^2 = 9B\Omega_0
\end{equation}
with the equilibrium atomic volume $\Omega_0$.

The mixed function $V_{\mathrm{AlLi}}$ was defined as a combination of a generalized Morse potential
and the repulsive terms also applied for the pure elments
\begin{multline}
    V_{\mathrm{AlLi}}(r) = [ \frac{D_0}{S-1}\exp \left( -\beta \sqrt{2S} (r-r_0) \right) - \\
    \frac{D_0S}{S-1}\exp \left( -\beta \sqrt{2/S} (r-r_0) \right) + \delta ] \Psi\left(\frac{r-r_c}{h}\right) \\
    +\sum_{n=1}^3 R(r, r_s^{(n)}, S^{(n)}).
\end{multline}
In the case of $\rho_{\mathrm{AlLi}}$ and $\rho_{\mathrm{LiAl}}$ the same function as for the single elements was applied.

\section{Selection of data}

\section{ACE potential configuration}
ACE basis configuration is provided in Table~\ref{tab:aceconfig}.
\begin{table}[tb!]
    \centering
    \caption{ACE basis configurations: cutoff radius ($r_c$), type of radial basis functions, $\nu$-order, n$_\mathrm{max}$, l$_\mathrm{max}$, and the maximum number of functions per element (\# func/elem) for each order $\nu$ for this configuration.}
    \begin{tabular}{cc}
    \hline
    r$_c$ &  7\,\AA \\
    Radial basis function &  SBessel \\
    $\nu$-order &  1/2/3/4/5/6 \\
    n$_\mathrm{max}$ &  15/6/4/3/2/2 \\
    
    l$_\mathrm{max}$ &  0/3/3/2/2/1 \\
    \# func/elem & 27/207/433/237/78/18  \\ 
    \hline
    
    \end{tabular}
    \label{tab:aceconfig}
\end{table}

\section{Settings of the RuNNer HDNNP fit}

\begin{table}[tbh]
    \centering
    \caption[Settings of the RuNNer HDNNP fit.]{Settings of the RuNNer HDNNP fit. Keywords and their values are listed as they are specified in the RuNNer input\.nn file format. ACSF settings are not included.}
    \begin{tabular}{ll}
        \toprule
        Keyword                         & Setting \\
        \midrule
        bond\_threshold                 & 0.5 \\
        calculate\_forces               & true \\
        center\_symmetry\_functions     & true \\
        cutoff\_type                    & 1 \\
        elements                        & Li Al \\
        epochs                          & 100 \\
        force\_update\_scaling          & 1.0 \\
        global\_activation\_short       & t t t l \\
        global\_hidden\_layers\_short   & 3 \\
        global\_nodes\_short            & 25 20 15 \\
        kalman\_lambda\_short           & 0.988 \\
        kalman\_nue\_short              & 0.9987 \\
        mix\_all\_points                & true \\
        nguyen\_widrow\_weights\_short  & true \\
        nn\_type\_short                 & 1 \\
        number\_of\_elements            & 2 \\
        optmode\_short\_energy          & 1 \\
        optmode\_short\_force           & 1 \\
        precondition\_weights           & true \\
        random\_seed                    & 90 \\
        repeated\_energy\_update        & true \\
        scale\_symmetry\_functions      & true \\
        short\_energy\_error\_threshold & 0.1 \\
        short\_energy\_fraction         & 1.0 \\
        short\_force\_error\_threshold  & 1.0 \\
        short\_force\_fraction          & 0.1 \\
        test\_fraction                  & 0.1 \\
        use\_short\_forces              & true \\
        use\_short\_nn                  & true \\
        write\_weights\_epoch           & 1 \\
        \bottomrule
    \end{tabular}
    \label{tab:hdnnp_settings_hyperparameters}
\end{table}

All HDNNPs mentioned in this work were generated using \textsc{RuNNer} (version 1.3), compiled with \textsc{ifort} (version 20230609) and linked against the MKL library (version 2023.2.0) as available through the \textsc{Intel oneAPI}. In addition to \textsc{pyiron}, \textsc{runnerase} (version 1.2.0) was used to facilitate calculation setup and evaluation.
The hyperparameters and settings that were chosen for training the HDNNP are given in Supplementary Tab.\,\ref{tab:hdnnp_settings_hyperparameters}. The employed atom-centered symmetry functions (ACSFs) were generated automatically using \textsc{runnerase} within a short-range cutoff radius of $R_{\mathrm{c}}=\SI{12.0}{\bohr}$. Nine radial ACSFs were generated for each of the four element combinations of the binary system with the hyperparameter $\eta=\SI{0.9}{\per\bohr\squared}$ and $R_{\mathrm{s}}$ spaced equally between the minimum pairwise distance of the given element and the cutoff radius. Additionally, multiple groups of angular ACSFs were created, permuting $\lambda=[1.0, -1.0]$, $\zeta=[1, 2, 4, 8, 16, 32, 64, 128, 256, 512]$ and four $\eta$ values per element triplet. In total, this yielded 258 ACSFs per element.

\section{$\mathrm{Al}_3\mathrm{Li}$ in the phase diagram}

\begin{figure*}
    \centering
     \includegraphics[width=0.8\textwidth]{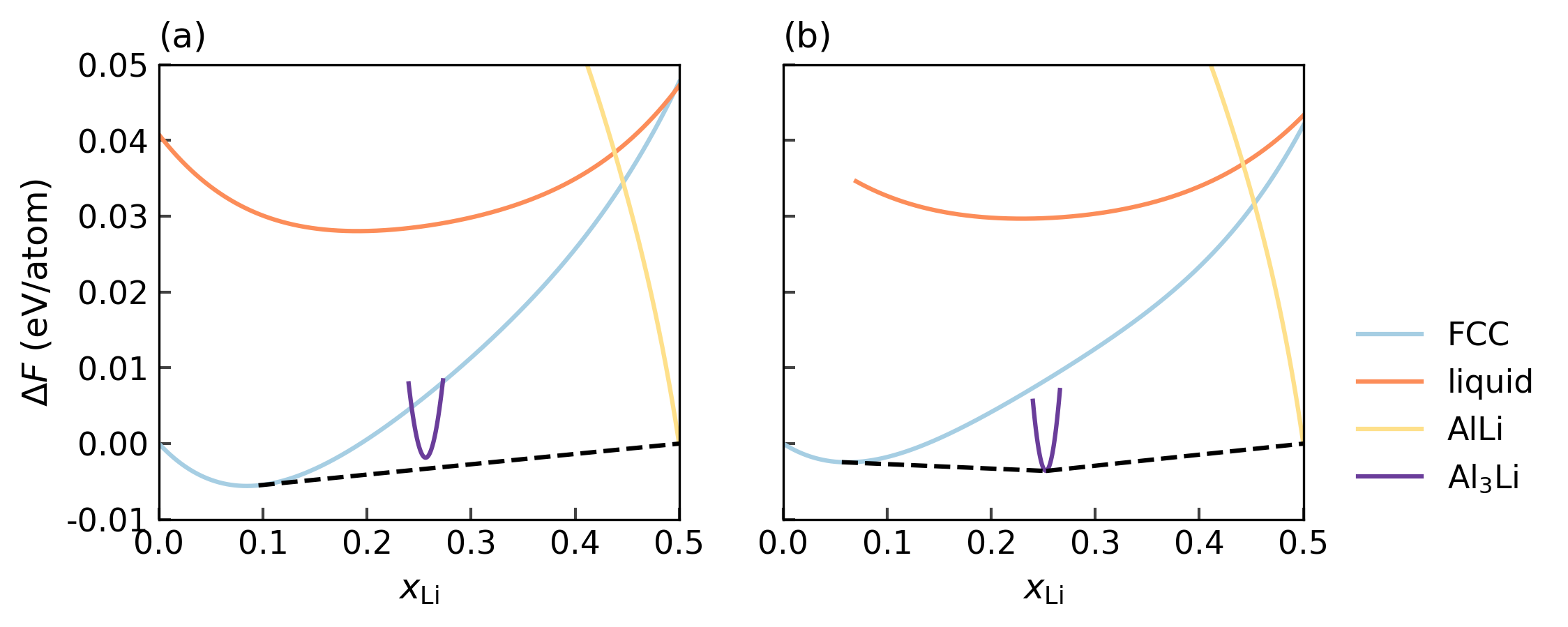}
    \caption{Free energy curves as a function of composition at 500 K for the (a) HDNNP and (b) ACE potential. $\mathrm{Al}_3\mathrm{Li}$ phase is above the common tangent that connects FCC Al and AlLi for the HDNNP potential, while it appears on the phase diagram for the ACE potential. }
    \label{fig:al3li_sup}%
\end{figure*}

Although the $\mathrm{Al}_3\mathrm{Li}$ appears on the DFT convex hull, it does not appear as a stable phase neither on the phase diagram \cite{Hallstedt2007}, nor on the CALPHAD phase diagram. The ACE potential predicts a fcc $+ \mathrm{Al}_3\mathrm{Li}$ and $\mathrm{Al}_3\mathrm{Li} +$ AlLi region which disappears at 580 K. For example, in the free energy curves at 500K, shown in Fig. \ref{fig:al3li_sup}, the ACE potential predicts a stable $\mathrm{Al}_3\mathrm{Li}$ phase that appears in the phase diagram. 
The HDNNP potential does not predict a stable $\mathrm{Al}_3\mathrm{Li}$ region in the investigated temperature range, although we note that this phase has been observed in a temperature range similar to ACE with an earlier HDNNP version exhibiting larger errors when employing different hyperparameters and data set filters.
Overall, it should be stressed that the subtle energy differences resulting in the emergence of this phase are in the order of the convergence level of the DFT calculations and the accuracy of the employed exchange correlation functional.

\section{Software availability}

The various software used in this work, along with their repository and documentation in shown in Table \ref{tab:software_availability}.

\begin{table*}[h!]
    \centering
    \caption{Repository and documentation of the various software tools employed in this work.}
    \begin{tabular}{lll}
        \toprule
        Software                        & Repository   & Documentation\\
        \midrule
        \texttt{pyiron}                 & https://github.com/pyiron/pyiron & https://pyiron.org/\\
        \texttt{pacemaker}              & https://github.com/ICAMS/python-ace & https://pacemaker.readthedocs.io/\\
        \texttt{RuNNer}                 & https://gitlab.com/runner-suite/runnerase & https://runner-suite.gitlab.io/\\
        \texttt{atomicrex}              & https://gitlab.com/atomicrex/atomicrex & https://atomicrex.org/\\
        \texttt{CALPHY}                 & https://github.com/ICAMS/calphy & https://calphy.org/\\
        \texttt{LAMMPS}                 & https://github.com/lammps/lammps & https://www.lammps.org/\\
        \texttt{pycalphad}              & https://github.com/pycalphad/pycalphad & https://pycalphad.org/\\
        \texttt{pyXtal}                 & https://github.com/qzhu2017/PyXtal & https://pyxtal.readthedocs.io/\\
        \bottomrule
    \end{tabular}
    \label{tab:software_availability}
\end{table*}

\bibliography{references}